\documentclass[12pt]{article}
\usepackage{amsmath}
\usepackage{amsthm}
\usepackage{amssymb}
\usepackage{graphics}
\usepackage{times}
\newtheorem{example}{Example}
\newtheorem{theorem}{Theorem}
\newtheorem{proposition}{Proposition}
\newtheorem{definition}{Definition}
\newtheorem{lemma}{Lemma}

\begin{document}
\thispagestyle{empty}
\noindent   \vspace*{30pt}\begin{center}
{\LARGE\textbf{Space-time and $\mathbb{G}_2$}\bigskip\vspace{10pt}\\
\textmd{Boris Doubrov,\\ \vspace{1pt}
Belarusian State University, \\ \vspace{1pt}Minsk, Belarus, \\\vspace{3pt}Jonathan Holland and George Sparling\\\vspace{1pt}Laboratory of Axiomatics,\\\vspace{3pt}
University of Pittsburgh,\\\vspace{8pt} Pittsburgh, Pennsylvania, USA}}\vspace{10pt}\\
\vspace{50pt}
{\large\textbf{Abstract}}\\\vspace{5pt}
\begin{quote}\textbf{A Weyl structure is a bundle over space-time, whose fiber at each space-time point is a space of maximally isotropic complex tangent planes. We develop the theory of Weyl connections for Weyl structures and show that the requirement that the connection be torsion-free fixes the  Weyl connection uniquely.  Further we show that to each such Weyl connection, there is naturally associated a (2, 3, 5)-Pfaffian system, as first analyzed by Cartan.  We determine the associated $\mathbb{G}_2$-conformal structure and calculate it explicitly in the cases of the Kapadia family of space-times and of the Schwarzschild solution.}
\end{quote}
\end{center}
\eject\noindent
\section*{Introduction} 
A space-time is by definition a connected smooth oriented four manifold, equipped with a Lorentzian metric.  Naturally associated to the metric of the space-time are a number of  Lie groups, for example the Lorentz group, the Poincare group, the Weyl group and the conformal group: these give rise to principal bundles over the space-time, with connection, which encode certain features of the space-time: for example the bundle of orthonormal frames carries the Levi-Civita connection,   whereas the Cartan normal conformal connection is carried by a principal bundle with group the Weyl group; this latter connection reflects the conformally invariant information of the space-time.  These structures easily generalise, mutatis mutandis, to other dimensions.\\\\The purpose of this work is to present a new space-time construction, which appears to be special to four dimensions, and which is conformally invariant, like the Cartan connection, but one for which the natural group is the first exceptional complex Lie group $\mathbb{G}_2$ of Wilhelm Killing \cite{kil}.  From a modern perspective, Killing found the $\mathbb{G}_2$ Lie algebra.  Later \'{E}lie Cartan and Fritz Engel gave realizations of the associated  Lie group, as symmetries of the octonions and as symmetries of a generic three-form in seven dimensions \cite{agr, bae, car1, eng}.  A priori there is no sign of this group in standard general relativity.  Apparently,   the reason for this is that the group does not appear in connection with Minkowski space-time:  instead the "flat model"  turns out to be a constant curvature null plane wave, whose conformal metric (a special case of a family of metrics considered by Devendra Kapadia) may be given, for example, as $dudv - dx^2 - u^2 dy^2$, where $(u, v, x, y)$ range over the open subset of $\mathbb{R}^4$ with $u$ positive \cite{kap}.\\\\
Cartan discovered that a $\mathbb{G}_2$-structure appeared automatically in connection with the study of generic Pfaffian systems consisting of three one-forms in five dimensions \cite{car2, dou1, dou2}.  Here we call this structure a $(2, 3, 5)$-system: the "$2$" refers to the space of vector fields that annihilate the Pfaffian system, the "$3$" to the extra direction generated by taking the Lie brackets of the vector fields of the ''$2$".  Then the "$5$'' indicates that the Lie brackets of the vector fields of the ''$3$'' taken amongst themselves generate the remaining two directions in the tangent-space.  
\eject \noindent  Given a real analytic space-time, it may be complexified and then it may be shown that its conformal structure gives rise to an holomorphic Riemann sphere bundle over the complexified space-time, which carries a natural conformally invariant $(2, 3, 5)$-structure, exactly of the type considered by Cartan, provided only that the space-time be not conformally flat.  In the following we develop this structure.  \\\\First we give a slight refinement of a conformal four-manifold, which we call a Weyl structure (which may be defined in any even dimension). This assigns to the conformal structure one of its two families of maximal isotropic subspaces:  in the language of physics, the Weyl structure is chiral, or parity-violating.  This family provides the relevant sphere bundle; on it, we define an extension of the Weyl structure, which we call a Weyl connection.  The Weyl connection has a naturally defined torsion and we show in the torsion-free case, the Weyl connection is unique.  When the Weyl torsion vanishes, the curvature is well-defined and is shown to be equivalent to the self-dual part of the Weyl curvature of the conformal manifold.  Finally, we show that if this curvature is non-zero, the torsion-free Weyl connection naturally gives a $(2, 3, 5)$-structure, which is therefore canonically associated to the conformal space-time.  In fact, even if the Weyl torsion is non-zero, generically, we still have a $(2, 3,5)$-structure, but the study of the structure in that case is far more complicated and has yet to be accomplished in detail.\\\\
Pawel Nurowksi, in completing a research program, begun with the third author, showed that naturally associated to Cartan's Pfaffian system, when put in a certain canonical form,  there is canonically defined a conformal structure on the underlying five-manifold of the system, whose Cartan conformal connection has group reducible to $\mathbb{G}_2$ and which gives the structure found by Cartan \cite{nur1, nur2, nur3}. Andreas \v{C}ap and Katja Sagerschnig showed how to describe the conformal structure directly from the Pfaffian system, without first putting it into canonical form \cite{cap}.  \\\\The main result of the present work is the determination of an explicit formula for the $\mathbb{G}_2$-conformal metric in five dimensions, for the case of torsion-free Weyl connections, using the work of \v{C}ap and Sagerschnig.  We also write out this conformal structure in two special cases:  first for the Kapadia family of plane wave metrics, where in particular, we identify "flat" models; then for the standard Schwarzschild metric \cite{schw}. For these metrics, we are able to put the system explicitly  into the Cartan canonical form and thereby compute the conformal structure, using Nurowski's formula.  For each case the two calculations of the conformal structure are carried out using different techniques and the results precisely agree.
\eject\noindent
The present work should be considered in the context of twistor theory; indeed one of the main achievements of twistor theory is the analysis of (complex analytic) space-times with vanishing self-dual Weyl curvature, due to Roger Penrose, Ezra Newman and others \cite{new, pen0, spa1, spa2}.  This work aims to fill the gap where the self-dual Weyl curvature is non-vanishing.  It would be of considerable interest to study manifolds, with non-zero self-dual Weyl curvature, but zero anti-self-dual Weyl curvature: these have associated $\mathbb{G}_2$-conformal structures and, at the same time, the dual twistor theory applies; the interaction between these theories should be fruitful.  Also even when the self-dual Weyl curvature is non-zero,  there is a twistor theory defined for each hypersurface in the space-time, studied particularly by the third author \cite{spa1, spa2}.  The interplay between the $\mathbb{G}_2$-theory and the hypersurface twistor theory has yet to be elucidated. \\\\We have presented the work for the case of the complexification of real space-times, which of course entails the physically undesirable requirement that the space-time be real analytic.  However the structure is still well-defined without analyticity and without complexification: one still has a sphere bundle and the $\mathbb{G}_2$-conformal structure associated with it, but the conformal structure is inherently complex, in that the sphere is still treated as if it had just one dimension (i.e. all calculations are holomorphic on the sphere), so it is a little hard to interpret the $\mathbb{G}_2$-structure in this case.  None of these difficulties occur in the real analytic case.  The $\mathbb{G}_2$-conformal structure is also present for Riemannian four-manifolds, where similar remarks apply.  In the case of Kleinian (ultra-hyperbolic) four-manifolds, however, the theory has a completely real version; in that case the group  is the real non-compact form of the complex Lie group $\mathbb{G}_2$ and analyticity is unneeded.\\\\
From our formula for the $\mathbb{G}_2$-conformal structure, it is possible to compute the curvature directly.   The results will be presented elsewhere.  The $\mathbb{G}_2$-conformal structure itself probes four derivatives of the space-time metric (two symmetrized derivatives of its Weyl curvature).  So the $\mathbb{G}_2$-conformal connection involves seven derivatives of the original space-time metric.  We expect that the structure will simplify, when we pass from the five-manifold to the associated seven-manifold of Charles Fefferman and C. Robin Graham \cite{fef1, fef2}.   In this context we should note that for the Kapadia family of metrics, the $\mathbb{G}_2$-conformal structure is always conformal to vacuum, so the Fefferman-Graham extension is routine.  At the time of writing, we do not yet know if this is true for the Schwarzschild $\mathbb{G}_2$-conformal structure.
\eject\noindent
In section one we recall the theory of $(2, 3, 5)$-structures, due to Cartan.  In section two, we define Weyl structures associated to a conformal structure for a four-dimensional vector space.  In section three we define Weyl space-times, which are equipped with Weyl structures for each tangent space.    In section four we construct the associated sphere bundle, which will carry the $\mathbb{G}_2$-conformal structure.  In section five we define Weyl connections, their torsion and curvature.  In section six, we compute the Weyl connection locally, and show that the torsion-free Weyl connection is unique.  In section seven, we show that every torsion-free Weyl connection with non-vanishing Weyl curvature is naturally a $(2, 3, 5)$-system.  
\\\\In section eight we recap the spinor approach to space-time following the approach of Penrose \cite{pen1,pen2}.  In section nine, we recall how to pass from a null tetrad in space-time to the associated spin connection.  In section ten we describe the abstract index approach of Penrose and in section eleven we apply this approach to the decomposition of the curvature tensor of a spin connection.  In section twelve we lift the spin connection to the spin bundle, constructing the appropriate invariant forms, which are dual to the horizontal vector fields of the connection.  In section thirteen we apply the spinor formalism to the Weyl structures and prove, in particular that the Weyl curvature of the Weyl connection coincides with the self-dual part of the Weyl curvature of the space-time conformal structure.  In section fourteen we use the spinor techniques to write out our main result, Theorem 2, the formula for the $\mathbb{G}_2$-conformal structure of a torsion-free Weyl connection.   The proof of  Theorem 2 occupies sections fifteen to twenty.
\\\\In section twenty-one, we introduce the Kapadia family of space-times.  In section twenty-two, we put the Weyl connection for each Kapadia space-time in the Cartan canonical form. In section twenty-three, we write out the $\mathbb{G}_2$-conformal structure for the Kapadia metrics, using the Nurowski formula and show that these agree with our general formula. We find that all of these $\mathbb{G}_2$-conformal structures are conformal to vacuum.   In section twenty-four, we specialize to a sub-family of the Kapadia family, where we are able to identify the elements of the sub-family, which give rise to conformally flat $\mathbb{G}_2$-conformal structures.  In section twenty-five, we recall the Schwarzschild solution and its spin connection. In section twenty-six, we write out the $\mathbb{G}_2$-conformal structure for Schwarzschild, which is surprisingly complicated, considering that it only depends on a single free parameter, the mass.  Finally, in section twenty-seven, we put the Schwarzschild Weyl connection in Cartan canonical form, compute the $\mathbb{G}_2$-conformal structure using Nurowski's formula and verify that it agrees with our general formula.
\eject\noindent
We end with some technical remarks:  in calculating with the Nurowski formalism, the calculations were carried out using the Maple computing system.  In computing the various spin connections and in deriving our formula for the \v{C}ap-Sagerschnig conformal structures, all the (intricate) calculations were done by hand, effectively following their calculation line by line.  Although we have only presented the results in the complex analytic case, all the metrics have obvious analogues in the Lorentzian, Riemannian and Kleinian real cases.  We have discussed the Penrose spinor structures in such a way that the natural group for these structures is the product of two copies of the Lie group $\mathbb{GL}(2, \mathbb{C})$; it is common to simplify by requiring that their structure group be reducible to the sub-group of all pairs $(A, B)$ in  $\mathbb{GL}(2, \mathbb{C})\times \mathbb{GL}(2, \mathbb{C})$, such that $\det(A) = \det(B)$.  We did not do this here.\\\\
We work in the holomorphic category; our manifolds are complex analytic and bundles over them are complex analytic.  If $\mathbb{M}$ is a complex manifold, and if if $k$ is a non-negative integer, we denote by $\Omega_k(\mathbb{M})$ its sheaf of holomorphic $k$-forms and by  $\Omega(\mathbb{M})$ the sheaf of all holomorphic forms on $\mathbb{M}$. The holomorphic tangent and cotangent bundles of $\mathbb{M}$ are denoted $\mathbb{TM}$ and $\mathbb{T^*M}$, respectively. We sometimes do not distinguish clearly between elements of a bundle or sheaf (germ) at a point and sections of the bundle or sheaf, over an open set, letting the context decide the appropriate interpretation. \\\\
If $\mathbb{A}$ is a (complex) vector space, or a vector bundle, we denote by $\mathbb{A}^*$ its dual; if $k$ is a non-negative integer, we denote by $\Omega^k(\mathbb{A})$ and $\mathbb{A}^k$, the $k$-th exterior and $k$-th symmetric products, of  $\mathbb{A}$ with itself, respectively.   Also we denote by $\Omega(\mathbb{A})$, the full exterior algebra of $\mathbb{A}$.  We denote by $\mathbb{PA}$ the projective space of $\mathbb{A}$, so $\mathbb{PA}$ is the Grassmanian of all one-dimensional subspaces of $\mathbb{A}$.  For $0 \ne a\in \mathbb{A}$ denote by $[a] \in \mathbb{PA}$ the one-dimensional subspace passing through $a$; so if $b \in \mathbb{A}$, then $b \in [a] $, if and only if a complex number $s$ exists with $b = sa$.   If $\mathbb{B}$ is a subspace of $\mathbb{A}$, we call the annihilator of $\mathbb{B}$, the subspace of $\mathbb{A}^*$, consisting of all $\beta\in \mathbb{A}^*$, such that $\beta(b) = 0$, for all $b \in \mathbb{B}$.  Our usual interpretation of a connection for a vector-bundle over a manifold is as a map $d$ from sections of the bundle to a section of the tensor product of the bundle with the bundle of one-forms of the manifold; it is understood that $d$ preserves duality of bundles; it is also understood that $d$ acts as the exterior derivative on forms.  Then the curvature of the connection is $d^2$.  Finally, when the context is appropriate, following Cartan, we often omit the wedge in the exterior product of forms.
\eject\noindent
\section{The five-variable theory of \'{E}lie Cartan}
Let $\mathcal{S}$ be a complex manifold of five dimensions and let $\mathcal{T}$ denote a two-complex dimensional sub-bundle of the tangent bundle of $\mathcal{S}$. Dually, let $\mathcal{T}'$ denote the three-dimensional subbundle of the co-tangent bundle of $\mathcal{S}$, that annihilates $\mathcal{T}$.  Then $\mathcal{T}$ is said to be generic, or of type $(2, 3, 5)$,  if, in the neighbourhood of any point of the space $\mathcal{S}$, there are holomorphic local sections $v$ and $w$ of $\mathcal{T}$,  such that the five vector fields $\{v, w, [v, w], [v, [v, w]], [w, [v, w]]\}$ trivialize the tangent bundle.
\\\\Following Gaspard Monge and David Hilbert, \'Elie Cartan analyzed such a system \cite{car2, hil, mon}.   Written first, dually, using differential forms, the bundle $\mathcal{T}'$ of Cartan may be taken to be the sub-bundle of the co-tangent bundle of a space $\mathcal{S}$, with five complex co-ordinates $(x, y, p, q, z)$, spanned by the following one-forms:
\[ dy - pdx, \hspace{10pt} dp- qdx, \hspace{10pt} dz - F(x, y, p, q, z) dx.\]
Here $F(x, y, p, q, z) $ is a given holomorphic function of its arguments.  The point here is that the vanishing of these forms (with $dx \ne 0$) corresponds to solutions of the differential system, where $'$ denotes differentiation with respect to $x$:
\[ y' = p, \hspace{10pt} p' = q, \hspace{10pt} z' = F(x, y, p, q, z).\]
Equivalently the system describes a single under-determined equation, studied, in special cases, by Monge and Hilbert:
\[ z' = F(x, y, y', y'', z).\]
For this system, the bundle $\mathcal{T}$ is then spanned by the vector fields:
\[ v = \partial_q, \hspace{10pt} w = \partial_x + p\partial_y + q\partial_p + F \partial_z, .\]
The required commutators are as follows:
\[ [v, w] = \partial_p + F_q \partial_z, \]
\[ [v, [v, w]] =  F_{qq} \partial_z,\]
\[ [w, [v, w]] =    - \partial_y  - (F_p + F_q F_z - F_{qx} - p F_{qy} - qF_{qp} - F F_{qz}) \partial_z.\]
Here and in the following we use subscripts to denote partial derivatives.  So we have, immediately, by inspection of these commutators:
\begin{lemma}
The Cartan system: $\mathcal{T}' = \{ dy - pdx, \hspace{10pt} dp- qdx, \hspace{10pt} dz - F(x, y, p, q, z) dx\}$ is generic, of type $(2, 3, 5)$, if and only if $F_{qq} \ne 0$.
\end{lemma}
\eject\noindent
Cartan studied the equivalence problem for his differential system and showed that in the generic case  it was governed by a principal bundle with connection with group $\mathbb{G}_2$ \cite{car2}.  Pawel Nurowski showed that the Cartan principal bundle could be interpreted as a reduction of the Cartan conformal connection for a conformal structure naturally defined on the space $\mathcal{S}$ \cite{nur1, nur2}.  Andreas \v{C}ap and Katja Sagerschnig showed a direct method of passing to the conformal structure associated to any $\mathcal{T}$ of type $(2, 3, 5)$, without requiring that the differential forms generating $\mathcal{T}'$ first be put in the form considered by Cartan \cite{cap}.  They also showed that, when applied to the Cartan $(2,3, 5)$-system, their conformal structure  agrees with that of Nurowski.  We may summarize with their theorem:
\begin{theorem}{(\'{E}lie Cartan, Pawel Nurowski, Andreas \v{C}ap and Katja Sagerschnig)}\\
To any $(2, 3, 5)$-system $\mathcal{T}$, on a five-dimensional complex manifold $\mathcal{S}$, there is naturally associated a conformal structure.  The Cartan conformal connection of this structure has holonomy a subgroup of $\mathbb{G}_2$.  
\end{theorem}
\noindent There is also a real version of this theorem, where all quantities are real; this requires only smoothness, not analyticity:  the conformal structure has signature $(3, 2)$ and the holonomy group of the Cartan conformal connection is then a subgroup of the non-compact real form of $\mathbb{G}_2$.
\eject\noindent
\section{Weyl structures}
Given a complex four-dimensional vector space, $\mathbb{T}$, denote by $\mathbb{PT}$ its associated projective space and by $\mathbb{GT}$ the Grassmanian of all two-dimensional subspaces of $\mathbb{T}$. The Klein quadric,  $\mathbb{KT}$, is the quadric in $\mathbb{P}(\Omega^2(\mathbb{T}))$, consisting of all $\omega \in \Omega^2(\mathbb{T})$, such that $\omega\wedge \omega = 0$. The Klein correspondence maps each $x \in \mathbb{GT}$ to the unique point $[\omega] $ of $\mathbb{KT}$, such that $v\wedge \lambda =0$, for any $v \in x$ and any $\lambda \in [\omega]$.  The inverse correspondence takes each  $[\omega]\in \mathbb{KT}$ to the unique element $x\in \mathbb{GT}$ consisting of all $v \in \mathbb{T}$, such that $v\wedge \lambda =0$,  for any $\lambda \in [\omega]$.
\\\\A projective plane $\Sigma$ in $\mathbb{P}(\Omega^2(\mathbb{T}))$, equivalently, a three-dimensional subspace of $\Omega^2(\mathbb{T})$,  is said to be regular, if and only if its intersection with $\mathbb{KT}$ is a non-singular conic.   If $\Sigma$ is regular, its polar plane, $\Sigma'$, is the space of all $\tau \in \Omega^2(\mathbb{T})$, such that $\sigma\wedge \tau = 0$, for all $\sigma \in \Sigma$.  Then $\Sigma'$ is regular and has polar $\Sigma$.  A pair of regular projective planes $\Sigma^\pm$, such that each is the polar of the other is called a polar pair.
\begin{definition}
A conformal structure for $\mathbb{T}$ is a non-singular quadric $[G]$ in $\mathbb{PT}$.   
\end{definition}
\begin{definition}  A two-dimensional subspace of $\mathbb{T}$ is said to be isotropic with respect to a conformal structure $[G]$ for $\mathbb{T}$,  if and only if the subspace is totally null, if and only if the projective image of the subspace is a projective line on the quadric $[G]$.
\end{definition}
\noindent  The isotropic planes in $\mathbb{T}$ assemble into two families, called the isotropic families, which each rule the quadric. Each family gives a projective curve in $\mathbb{KT}$; in turn each such curve in $\mathbb{KT}$ is the intersection with $\mathbb{KT}$ of a (unique) regular projective plane in  $\mathbb{P}(\Omega^2(\mathbb{T}))$.  The two planes thus generated form a polar pair.  Conversely given a polar pair of projective planes in $\mathbb{P}(\Omega^2(\mathbb{T}))$, each plane intersects the Klein quadric in a projective curve, giving the pair of isotropic families for a unique conformal structure for $\mathbb{T}$.  So we have the Lemma:
\begin{lemma}  There is a one-to-one correspondence between polar pairs of projective planes in $\Omega^2(\mathbb{T})$ and conformal structures for $\mathbb{T}$.
\end{lemma}
\noindent The annihilator of a regular plane in $\mathbb{P}(\Omega^2(\mathbb{T}))$ is a regular plane in $\mathbb{P}(\Omega^2(\mathbb{T}^*))$, which gives $\mathbb{T}^*$ the conformal structure inverse to that  of $\mathbb{T}$.  The annihilator of an isotropic plane in $\mathbb{T}$ is then an isotropic plane in $\mathbb{T}^*$.  Henceforth, we pass freely from $\mathbb{T}$ to $\mathbb{T}^*$ and back, via annihilators.
\begin{definition} A Weyl structure is a pair $(\mathbb{T}, \Sigma)$ consisting of  $\mathbb{T}$, a four-dimensional complex vector space and $\Sigma$, a regular three-dimensional subspace of $\Omega^2(\mathbb{T})$.\end{definition}
\noindent Thus a Weyl structure determines a conformal structure for $\mathbb{T}$ and a distinguished family of projective lines on the quadric $[G]$ defining the conformal structure. 
\eject\noindent
\section{Conformal space-times; Weyl space-times}
We work with complex manifolds, entirely in the holomorphic category.  For convenience, we restrict our attention to connected manifolds.
\begin{definition} A complex conformal space-time is a pair $(\mathbb{M}, [G])$ consisting of a complex four-manifold $\mathbb{M}$ and a holomorphic family, $[G] = \{[G]_x; x\in \mathbb{M}\}$, where $[G]_x$ is a conformal structure  for the tangent space to $\mathbb{M}$ at $x\in \mathbb{M}$. 
\end{definition}
\noindent By the results of the last section, the following definition is equivalent:
\begin{definition} A complex conformal space-time is a triple $(\mathbb{M}, \Sigma^\pm)$ consisting of a complex four-manifold $\mathbb{M}$ and a polar pair of sub-bundles, $\Sigma^\pm$ of $\Omega_2(\mathbb{M})$.
\end{definition}
\noindent For a Weyl space-time, we simply single out one of the elements of the polar pair:
\begin{definition}A Weyl space-time is a pair $(\mathbb{M}, \Sigma)$ consisting of a complex four-manifold and an everywhere regular three dimensional subbundle, $\Sigma$, of $\Omega_2(\mathbb{M})$.
\end{definition}
\noindent  So a Weyl space-time gives $\mathbb{M}$ a holomorphic family of Weyl structures, one for each tangent space of $\mathbb{M}$.  We denote by $[G]_\Sigma$ the conformal structure on $\mathbb{M}$, naturally derived from the Weyl manifold $(\mathbb{M}, \Sigma)$.   So every Weyl space-time is naturally a complex conformal space-time.  If $(\mathbb{M}, \Sigma^+)$ is a Weyl space-time, then so is $(\mathbb{M}, \Sigma^-)$, where $\Sigma^-$ is the polar of $\Sigma^+$.  Then we have $[G]_{\Sigma^+} = [G]_{\Sigma^-}$.
\begin{example} The Klein quadric
\end{example} 
\noindent Let $\mathbb{T}$  be a three-dimensional complex projective space.  Denote by $\mathbb{M}$ the space of projective lines in $\mathbb{T}$.   So $\mathbb{M}$ is a four-manifold, the Klein quadric of $\mathbb{T}$. Then if $x$ is a projective line in $\mathbb{T}$, so $x$ is a point of $\mathbb{M}$, there is a three-dimensional cone at $x$, the space of all projective lines $y$ in $\mathcal{T}$ that pass through the line $x$.  To each such line $y \ne x$, there is naturally associated its point of incidence $x\cap y$ with $x$ and the plane containing $x$ and $y$, $x\cup y$.  The map $y \rightarrow (x\cap y, x\cup y)$ fibers the null cone over the quadric $[G]_x$ of all pairs $(z, Z)$ consisting of points $z\in \mathcal{T}$ of $x$ and planes $Z$ in $\mathcal{T}$ such that $x \subset Z$.  As $x$ varies, the ensemble of quadrics $\{[G]_x: x\in \mathcal{T}\}$ represents a conformal structure for $\mathbb{M}$.  At each $x\in \mathbb{M}$, fixing $z$ and letting $Z$ vary gives one Weyl structure, $\Sigma^+$ , say;  fixing $Z$ and letting $z$ vary gives the other Weyl structure, say $\Sigma^-$. 
\eject\noindent
\section{The null cone bundle of a Weyl space-time}
The null cone bundle of  a Weyl space-time $(\mathbb{M}, \Sigma)$  is the fiber bundle $\mathbb{N}$ consisting of all $\omega \in \Sigma$ such that $\omega\wedge \omega = 0$.  Projectively this gives a fiber bundle, denoted, $\mathbb{S}$, over $\mathbb{M}$, with fibre at $x \in \mathbb{M}$, a Riemann sphere, denoted $\mathbb{S}_x$. In particular, $\mathbb{S}$ is a five-dimensional complex manifold.  \\\\Denote by $p$ the canonical projection $p: \mathbb{S} \rightarrow \mathbb{M}$.  Denote by $p^*$ the induced map from the tangent bundle, $\mathbb{TS}$, of $\mathbb{S}$, to the tangent bundle of $\mathbb{M}$.  Denote by $\mathbb{V}$ the kernel of $p^*$, so $\mathbb{V}$ is a line sub-bundle of $\mathbb{TS}$, the vertical bundle. The elements of $\mathbb{V}$ are tangent to the fibers of $p$.  Restricted to any fiber, $\mathbb{S}_x$ for $x \in \mathbb{M}$, $\mathbb{V}$ gives a line bundle of Chern class two. On each fiber, the three-dimensional space of global sections of $\mathbb{V}$ forms the Lie algebra of $\mathbb{O}(3, \mathbb{C})$, under commutation.\\\\  If $X = (x, [\omega])\in \mathbb{S}$ (so we have, in particular, $p(X) = x \in \mathbb{M}$, whereas $0 \ne \omega \in \Sigma$, where $\omega\wedge \omega = 0$), denote by $\mathbb{W}_X$, the space of all tangent vectors $Y$ at $X$, such that if $y = p^*(Y)$ is the projected tangent vector at $x$, then $\iota_y(\lambda) = 0$, for any $\lambda \in [\omega]$.  Then $\mathbb{W} = \{\mathbb{W}_X; X\in \mathbb{S}\}$ is a three-dimensional vector sub-bundle of $\mathbb{TS}$; we call $\mathbb{W}$ the Weyl bundle of $\mathbb{S}$.   Note that $\mathbb{V} \subset \mathbb{W}$.  Restricted to any fiber, $\mathbb{S}_x$ for $x \in \mathbb{M}$, the bundle $\mathbb{W}$ splits (following the Birkhoff theorem) as a sum of three line bundles with Chern classes $(2, 1, 1)$.  In particular the space of splittings of the inclusion homomorphism $\mathbb{V} \rightarrow \mathbb{W}$ on any $\mathbb{S}_x$ is parametrized by the space of global sections of a bundle with Chern classes $(1, 1)$, so is a four-dimensional vector space. 
\\\\
Dually, for $X \in \mathbb{S}$, denote by $\Theta_{X}$, the space of all one-forms $\alpha$ at $p(X)$, such that  $\alpha\wedge \lambda =0$, for any $\lambda \in [\omega]$, pulled back to the point $X$ along the canonical projection.  Then as $X$ varies,  $\Theta = \{\Theta_X; X \in \mathbb{S}\}$ gives a two-dimensional subbundle of $\Omega_1(\mathbb{S})$.  Then the bundles $\Theta$ and $\mathbb{W}$ are the annihilators of each other.   Next, let $\mathbb{V}'$ denote the annihilator of $\mathbb{V}$ in $\Omega_1(\mathbb{S})$.  So $\mathbb{V}'$ is a four-dimensional sub-bundle of $\Omega_1(\mathbb{M})$.  Also $\mathbb{V}'$ is the pull-back along the projection $p$ of $\Omega_1(\mathbb{M})$.  $\mathbb{V}'$ may be called the bundle of tensorial one-forms of $\mathbb{S}$.  Finally $\Theta \subset \mathbb{V}'$.
\begin{example} The Klein quadric
\end{example}   
\noindent For the Klein quadric, $\mathbb{M}$, of example one above, in the case of the Weyl structure $\Sigma^+$, the bundle $\mathbb{S}$ is the space of pairs $(x, z)$ with $x$ a projective line in $\mathbb{T}$ and $z \in x$; for the Weyl structure $\Sigma^-$, the bundle $\mathbb{S}$ is the space of pairs $(x, Z)$ with $x$ a projective line in $\mathbb{T}$ and $Z$ a projective plane in $\mathbb{T}$, such that  $x\in Z$.  
\eject\noindent
\section{Weyl connections, their torsion and curvature}
\noindent   Let $(\mathbb{M}, \Sigma)$ be a Weyl space-time, with its projective null cone bundle $\mathbb{S}$, the projection $p: \mathbb{S} \rightarrow \mathbb{M}$, the Weyl bundle $\mathbb{W} \subset \mathbb{TS}$, the vertical bundle $\mathbb{V} \subset \mathbb{W}$ and the bundle of one-forms, $\Theta$, the annihilator of $\mathbb{W}$, as described in the last section.
\begin{definition} A Weyl connection is a two-dimensional subbundle $\mathbb{T}$ of $\mathbb{W}$, such that $\mathbb{W} = \mathbb{T} + \mathbb{V}$.  Dually a Weyl connection is a three-dimensional sub-bundle $\mathbb{T}'$ of $\Omega_1(\mathbb{S})$ such that $\mathbb{T'}\cap \mathbb{V}' = \Theta$; we pass from $\mathbb{T}$ to $\mathbb{T}'$ and back via annihilators.
\end{definition}
\noindent On each fiber $\mathbb{S}_x$, $\mathbb{T}$ is a sum of  two line bundles, each of Chern class one.
Given a Weyl connection, $\mathbb{T}$, for $(\mathbb{M}, \Sigma)$, consider the bundle $\Omega_2(\mathbb{S})$ modulo the ideal generated by $\mathbb{T}'$.  On dimensional grounds, this is a line bundle, denoted $\mathbb{L}$, over $\mathbb{S}$.  Note that $\mathbb{L}$ may be identified with the quotient of the kernel of $\mathbb{V}$ in $\Omega_2(\mathbb{S})$ modulo the ideal generated by $\Theta$.   Also $\mathbb{L}$ has Chern class $2$ over each $\mathbb{S}_x$.
\begin{definition}For $\alpha$ any local section of the bundle $\Theta$, put $T(\alpha) = d\alpha$ mod $\mathbb{T'}$.  Then $T$ gives a vector bundle homomorphism from $\Theta$ to $\mathbb{L}$, called the Weyl torsion of the Weyl connection, $(\mathbb{M}, \Sigma, \mathbb{T})$.
\end{definition}
\begin{definition}A Weyl connection $(\mathcal{M}, \Sigma, \mathbb{T})$ is torsion-free if and only if $T =0$, if and only if $d\alpha$ lies in the ideal generated by $\mathbb{T}'$, for any local section $\alpha$ of $\Theta$.
\end{definition}
\noindent  Now suppose that $(\mathbb{M}, \Sigma, \mathbb{T})$  is a torsion-free Weyl connection.  For $\alpha$ a local section of $\mathbb{T}'$, put $W(\alpha) = d\alpha$ mod $\mathbb{T}'$.  Then $W$ is a vector bundle homomorphism from $\mathbb{T}'$ to $\mathbb{L}$ which vanishes on $\Theta$, so $W$ may be considered to be a homomorphism of line bundles $W: \mathbb{T}'/\Theta \rightarrow \mathbb{L}$.  Further, the line bundles $\mathbb{V}$ and $ \mathbb{T}'/\Theta $ are naturally dual, so $W$ may be considered to be a global section of the line bundle $\mathbb{L} \otimes \mathbb{V}$, a line bundle, which, over each fiber $\mathbb{S}_x$, has Chern class four.
\begin{definition}The global section $W$ of the line bundle $\mathbb{L} \otimes \mathbb{V}$ over $\mathbb{S}$ is called the Weyl curvature of the torsion-free Weyl connection $(\mathbb{M}, \Sigma, \mathbb{T})$. \end{definition}
\begin{definition}A Weyl connection $(\mathbb{M}, \Sigma, \mathbb{T})$ is said to be Weyl-flat, or a twistor structure, if and only if its Weyl torsion and Weyl curvature both vanish identically, if and only if $\mathbb{T}'$ defines a differential ideal, if and only if $\mathbb{T}$ is Frobenius integrable.
\end{definition}
\begin{definition} If $(\mathbb{M}, \Sigma, \mathbb{T})$ is a twistor structure, its space of integral manifolds is a three-dimensional space, called the twistor space of the Weyl manifold.\end{definition}
\begin{example}The Klein quadric
\end{example}  \noindent For the Klein quadric, $\mathbb{M}$, of examples one and two above, fixing $z$ and varying $x$ foliates $\mathbb{S}$ gives $(\mathbb{M}, \Sigma^+)$ a flat Weyl connection. Similarly, fixing $Z$ and varying $x$ gives $(\mathbb{M}, \Sigma^-)$ a flat Weyl connection.  In each case the twistor space is $\mathbb{T}$.  
\eject\noindent
\section{Local computations}
Let $(\mathbb{M}, \Sigma, \mathbb{T})$ be a Weyl manifold with connection.  A normalized frame for $\Sigma$ is a local basis $\{\sigma_\pm, \sigma_0\}$ such that $\sigma_\pm^2 = 0$, $\sigma_\pm \sigma_0 = 0$ and $\sigma_+\sigma_- =  - 2\sigma_0^2 =  \tau\ne 0$.  Such a normalized frame always exists.  With respect to a normalized frame, the general element $\sigma $ of $\Sigma$ can be written, uniquely $\sigma = x_+\sigma_+  + x_-\sigma_- + 2x_0\sigma_0$.  Then we have $\sigma^2 = 2(x_+x_- - x_0^2)\tau$.
In particular $\sigma$ lies in the null cone of $\Sigma$ if and only if $\sigma^2 =0$, if and only if $x_+x_- = x_0^2$, if and only if complex numbers $p$ and $q$ exist such that $\sigma = p^2\sigma_+ + 2pq\sigma_0 + q^2 \sigma_-$.  The ratios $p:q$ then parametrize the sphere bundle $\mathbb{S}$.
Next it is straightforward to show that a local basis of one-forms, $\{l, m, m', n\}$ exists, called a null tetrad, such that the symmetric tensor, $G = 2(ln - mm')$, represents the conformal structure of the Weyl manifold and such that $\sigma_+ = lm', \sigma_- = mn$ and $2\sigma_0 = ln + mm'$; then $\tau = -lmm'n$.  Note that the polar, $\Sigma'$, of $\Sigma$ has as its local basis $\{lm, m'n, ln - mm'\}$.  At any point $(x, p, q)$ of $\mathbb{S}$, put:
\[ \eta = pl + qm, \hspace{10pt}\theta = pm' + qn.\]
Then we have:
\[  p^2\sigma_+ + 2pq\sigma_0 + q^2 \sigma_- = p^2lm' + pq(ln + mm') + q^2 mn\]
\[ = (pl + qm)(pm' + qn) = \eta\theta.\]
Thus the pair of one-forms  $\{\eta,  \theta\}$ span an isotropic space for any $p$ and $q$, not both zero.   Also a local basis for the bundle $\Theta$ at $(x, p, q) \in \mathbb{S}$ is the pair of one-forms $\{\eta, \theta\}$.  If now $\mathbb{T}$ is a Weyl connection, the third basis form in $\mathbb{T}'$, apart from $\eta$ and $\theta$, can be written locally as $\gamma =qdp - pdq  - \Gamma_1 l - \Gamma_2 m -  \Gamma_3 m' - \Gamma_4 n $,  where each of the functions $\Gamma_1, \Gamma_2, \Gamma_3$ and $\Gamma_4$ is homogeneous of degree two in the pair $(p, q)$.  Now if we work modulo the ideal generated by $\Theta$, there are one-forms $\phi$ and $\psi$, such that we have: 
\[ l = - q\phi,  \hspace{10pt} m = p\phi, \hspace{10pt} m' = q\psi,  \hspace{10pt} n = - p\psi.\]
Put $\sigma = \phi\psi$.  Then modulo the ideal of $\Theta$, we have:
\[ lm' = -q^2\sigma, \hspace{10pt}lm = m'n = 0, \hspace{10pt} ln = mm' = pq\sigma, \hspace{10pt} mn = - p^2\sigma.\]
In particular, modulo $\Theta$, $\Sigma'$ is reduced to zero (i.e. $\Sigma'$ lies in the ideal).  Then we compute the exterior derivatives of the members of the null tetrad and reduce the results modulo the ideal of $\Theta$.  This gives formulas, valid for $A$, $B$, $C$ and $D$ certain (computable) homogeneous quadratic polynomials in the pair $(p, q)$:
\[ dl = A\sigma, \hspace{10pt}dm =  B\sigma,\hspace{10pt}dm' = C\sigma, \hspace{10pt}dn = D\sigma.\]
\eject\noindent
Now we may compute the torsion of $\mathbb{T}'$; working modulo the ideal of $\Theta$, we have first:
\[ d\eta = d(pl + qm) = (dp) l + (dq)m + p dl + qdm\]
\[ = (pdq-qdp)\phi + (pA + qB)\sigma\]
\[ = \phi\gamma -  ((p\Gamma_2 - \Gamma_1q) \phi  - (p\Gamma_4 - q\Gamma_3)\psi)\phi + (pA + qB)\sigma\]
\[ = \phi\gamma +  (p(A - \Gamma_4) + q(B + \Gamma_3))\sigma.\]
Since the first term lies in the ideal of $\mathbb{T}'$, we see that the first part of the torsion is represented by the quantity $\tau_1 = p(A - \Gamma_4) + q(B+\Gamma_3)$, which is homogeneous of degree three in $(p, q)$.  Next we have, modulo the ideal of $\Theta$:
\[ d\theta = d(pm' + qn) =  (dp)m' + (dq)n + (pC + qD) \sigma\]
\[ =  (qdp - pdq)\psi + (pC + qD) \sigma\]
\[ =  \gamma\psi  + ((p\Gamma_2 - \Gamma_1q) \phi  - (p\Gamma_4 - q\Gamma_3)\psi)\psi + (pC + qD) \sigma\]
\[ =  \gamma\psi  + ((p(C + \Gamma_2) + q(D -  \Gamma_1)) \sigma.\]
Since the first term lies in the ideal of $\mathbb{T}'$, the second and last part of the torsion is represented by the quantity $\tau_2 = p(C + \Gamma_2)+ q(D - \Gamma_1)$, which is also homogeneous of degree three in $(p, q)$.   \\\\Now suppose the torsion is zero: $\tau_1 = \tau_2 = 0$.  Then we can write out the forms $\Gamma_1, \Gamma_2, \Gamma_3$ and $\Gamma_4$ as follows:
\[ \Gamma_1 = D + p g, \hspace{10pt} \Gamma_2 = -C + qg, \hspace{10pt} \Gamma_3 = - B+ ph, \hspace{10pt}\Gamma_4 = A + q h.\]
Here $g$ and $h$ are unknown homogenous functions of degree one in $(p, q)$.  Then we have:
\[ \Gamma_1 l + \Gamma_2 m + \Gamma_3 m' + \Gamma_4 n =  Dl- Cm - Bm' +An  + g\eta + h\theta.\]
So, in the zero torsion case, the bundle $\mathbb{T'}$ has as basis the one-forms: 
\[ \mathbb{T'} = \{pl + qm,  pm' + qn, qdp - pdq - Dl + Cm + Bm' -An \}.\]
Thus we have the analogue of the Levi-Civita Lemma for Weyl manifolds:
\begin{lemma}(Levi-Civita for Weyl) Given the Weyl manifold, $(\mathbb{M}, \Sigma)$, there is a unique torsion-free Weyl connection $(\mathbb{M}, \Sigma, \mathbb{T}')$.
\end{lemma}
\eject\noindent
\section{Torsion-free Weyl is naturally (2, 3, 5)}
In the torsion-free case, to compute the Weyl curvature, using the notation of the last section, we need only compute the exterior derivative of the one-form $\gamma = qdp - pdq - Dl + Cm + Bm' -An$ modulo the ideal generated by $\mathbb{T'}$. The result necessarily has the form:
\[ d\gamma = W \sigma.\]
Here $W$ is a computable homogeneous quartic polynomial in the variables $(p, q)$ that represents the Weyl curvature.  \\\\Now we assume henceforth that $\mathbb{T}$ is \emph{not} integrable, so not a twistor space.  Hence $W $ is not identically zero.  We delete from the space $\mathbb{S}$ the zeroes of $W$.  This entails first deleting from $\mathbb{M}$ any point at which the Weyl curvature vanishes and the whole fiber of $\mathbb{S}$ at that point.  Denote by $\mathcal{M}$ the residual manifold, an open subset of $\mathbb{M}$, with complement in $\mathbb{M}$ an analytic set.  Next at any point $x \in \mathcal{M}$, since $W$ is a not identically zero quartic, by the fundamental theorem of algebra, there are at least one and at most four values of the ratio $p:q$, where $W$ vanishes.    We delete these points $(x, p, q)$ from $\mathbb{S}$.  Denote the residual space by $\mathcal{S}$.  This is an open subset of $p^{-1}(\mathcal{M})$, with closure $\mathbb{S}$, whose complement in $\mathbb{S}$ is an analytic set.  Denote by $\mathcal{T}$ and $\mathcal{T}'$, the restrictions of $\mathbb{T}$ and $\mathbb{T}'$ to $\mathcal{S}$.  Then we have the proposition:
\begin{proposition} The torsion-free Weyl connection $(\mathcal{M}, \mathcal{S}, \mathcal{T})$ is a $(2, 3, 5)$-system on the five manifold $\mathcal{S}$.
\end{proposition}
\noindent  Note that, in general, $\mathcal{S}$ is not globally a fiber bundle over $\mathcal{M}$, since the number of roots of $W$ may vary from point to point.  However in the generic case, where there is at least one point where $W$ has four distinct roots, then, perhaps after deleting a further algebraic set,  we may assume that $W$ has four distinct roots everywhere.  In that case, called algebraically general, it is natural to replace each sphere $\mathbb{S}_x$, for $x \in \mathcal{M}$, by its double cover, a torus, branched at the four roots.   The modulus of the torus is determined by the cross-ratio of the four points and is expressible directly in terms of invariants of $W$.  Then we have a toroidal fibration over $\mathcal{M}$ and the proposition implies that we have a $(2, 3, 5)$-structure away from the branch points.
\eject\noindent
For the proof of the proposition, we work locally and use the notation of the last section.  Denote the dual basis of vector fields on $\mathcal{M}$ by $\{ L, M, M', N\}$, dual to the basis of one-forms $\{l, m, m', n\}$, with the dualities $L.n = N.l = 1, M.m' = M'.m = -1$ and all other dot products zero.  Then the vector fields spanning $\mathcal{T}$ are the vector fields:
\[ \{  pL + qM - X\partial, \hspace{10pt} pM' + qN - Y \partial\}.\] 
Here these vector fields act on functions of $p$ and $q$ that are homogeneous of degree zero.   Then, acting on such a function $f$, the vertical operator $\partial$ is determined by the relations: $(\partial_p, \partial_q) f = (-q, p)\partial f$.  Note that $\partial$ is dual to the homogeneous one-form $pdq - qdp$: $(pdq - qdp).\partial = 1$.  Also $\partial$ is of degree minus two.  Then the product of $\partial$ and any homogeneous function of degree two in $(p, q)$ maps the space of homogeneous  functions of degree zero to itself.  The vector fields of $\mathcal{T}$ automatically annihilate the forms of $\Theta$; then the quantities $X$ and $Y$, each homogeneous of degree three in the variables $(p, q)$, must also annihilate the third basis one-form of $\mathcal{T}'$, the one-form  $\gamma =  qdp - pdq - Dl + Bm + Cm' - An$, which (since $ (pdq - qdp).\partial = 1)$ gives the relations:
\[ X = (pL + qM).(Dl- Cm - Bm' +An) = pA + qB.\]
\[ Y =   (pM' + qN).(Dl- Cm - Bm' +An) = pC + qD, \]
So we may rewrite the vector fields as:
\[ \{ P, Q\}, \hspace{10pt} P =  pL + qM  + B\partial_p - A \partial_q,\hspace{10pt} Q = pM' + qN + D\partial_p - C\partial_q.\]
In this form they act naturally on any function of $p$ and $q$, not necessarily just functions homogeneous of degree zero.  Note that the  product of $P$ or $Q$ by any homogeneous function of degree minus one in $(p, q)$ maps the space of homogeneous  functions of degree zero to itself.  Then, since the Weyl torsion is zero,  their Lie bracket, acting on functions of degree zero, considered modulo combinations of $P$ and $Q$, gives the relation:
\[ [P, Q] = W \partial, \mod P, Q.\]
Since, by hypothesis, the Weyl curvature $W$ is everywhere non-zero, we have that the Lie bracket adds to the $"2"$ of $P, Q$ the vertical vector field $\partial$ to give the $"(2, 3)"$, of our putative $(2, 3, 5)$ system. 
\eject\noindent
Now we have, acting on homogeneous functions of degree zero, when $p \ne 0$:
\[ [p^2\partial, p^{-1}P] = [p\partial_q, p^{-1}P] = p^{-1}[p\partial_q, P] =  M,  \mod P, Q, \partial, \]
\[ [p^2\partial, p^{-1}Q] = [p\partial_q, p^{-1}Q] = p^{-1} [p\partial_q, Q] = N, \mod P, Q, \partial.\]
So, when $ p\ne 0$ the span of the iterated commutators gives at least the span of the five vector fields:
\[ \{p^{-1}P, p^{-1} Q, p^2\partial, M, N\}.\]
This is equivalent to the span of the set $\{ p^2\partial, L + qp^{-1} M, M' + qp^{-1} N, M, N\}$ and therefore to the span of the set: $\{ p^2\partial, L, M, M', N\}$, which spans the entire tangent space.  Similarly when $q \ne 0$, we have:
\[ [- q^2\partial, q^{-1}P] = [q\partial_p, q^{-1}P] =   q^{-1}[q\partial_p, P] = L,   \mod P, Q, \partial, \]
\[ [-q^2\partial, q^{-1}Q] = [q\partial_p, q^{-1} Q] =  q^{-1}[q\partial_p, Q] =  M', \mod P, Q, \partial. \]
So, when $ q\ne 0$ the span of the iterated commutators gives at least the span of the five vector fields:
\[ \{q^{-1}P, q^{-1} Q, q^2\partial, L, M'\}.\]
This is equivalent to the span of the set $\{ q^2\partial, M + pq^{-1}L, N + pq^{-1}M', L, M'\}$ and therefore to the span of the set: $\{ q^2\partial, L, M, M', N\}$, which spans the entire tangent space.  Since the variables $p$ and $q$ are never both zero, we have proved that the system $\mathcal{T}$ is a $(2, 3, 5)$ system on the space  $\mathcal{S}$, as required, and we are done.  \\\\Note that we have only proved this result here for the torsion-free case; when suitably stated the result also holds true in the case of non-zero torsion, but this will not be analyzed here.
\eject\noindent
\section{Spinor computations}
We reprise the approach of Penrose to spinors for a complex four-manifold, $\mathbb{M}$ \cite{pen1, pen2}:
\begin{definition} A Penrose spin structure for $\mathbb{M}$ is a quartet of two-dimensional bundles over $\mathbb{M}$,  $\mathbb{S}^\pm, \mathbb{S}_\pm$, with $\mathbb{S}^+$ dual to $\mathbb{S}_+$ and $\mathbb{S}^-$ dual to $\mathbb{S}_-$, together with a one-form $\theta$ on $\mathbb{M}$ with values in $\mathbb{S}^+\otimes \mathbb{S}^-$, called the canonical one-form, subject to the regularity condition that $\theta$ give an isomorphism from $\mathbb{TM}$ to $\mathbb{S}^+ \otimes \mathbb{S}^-$; equivalently, $\theta$ is required to give an isomorphism from $\mathbb{S}_+ \otimes \mathbb{S}_-$ to $\mathbb{T^*M}$.
\end{definition}
\noindent Denote by $\mathbb{L}^\pm$ and $\mathbb{L}_\pm$,  the line bundles $\Omega^2(\mathbb{S}^\pm)$ and $\Omega^2(\mathbb{S}_\pm)$, respectively. When considering tensor products of the spin spaces, we ignore the relative ordering of factors from $\mathbb{S}^+$ or $\mathbb{S}_+$ vis \`{a} vis factors from $\mathbb{S}^-$ or $\mathbb{S}_-$: more formally, we quotient the tensor algebra at each point of $\mathbb{M}$, by the two-sided ideal generated by the relations $\alpha^+\otimes \alpha^- - \alpha^-\otimes \alpha^+$, for all $\alpha^+$ in $\mathbb{S}^+$, or in $\mathbb{S}_+$ and all $\alpha^-$ in $\mathbb{S}^-$, or in $\mathbb{S}_-$.   Similarly, we regard the line bundles $\mathbb{L}^\pm$ and $\mathbb{L}_\pm$ as commuting with other spinors.
\\\\
Taking appropriate tensor products, we have the decompositions:
\[  (\mathbb{TM})^2 =  \mathbb{L}^+\otimes \mathbb{L}^- + (\mathbb{S}^+)^2\otimes (\mathbb{S}^-)^2, \]
\[  (\mathbb{T^*M})^2 =  \mathbb{L}_+\otimes \mathbb{L}_- + \mathbb{S}_+^2\otimes \mathbb{S}_-^2, \]
\[ \Omega^2(\mathbb{TM}) = \Sigma^+  + \Sigma^{-}, \]
\[ \Sigma^+=  \mathbb{L}^+ \otimes (\mathbb{S}^-)^2,\hspace{10pt} \Sigma^- =  (\mathbb{S}^+)^2\otimes \mathbb{L}^-,\]
\[ \Omega_2(\mathbb{M}) = \Sigma_+  + \Sigma_{-}, \]
\[ \Sigma_+=  \mathbb{L}_+ \otimes \mathbb{S}_-^2,\hspace{10pt} \Sigma_- =  \mathbb{S}_+^2\otimes \mathbb{L}_-,\]
\[ \Omega^3(\mathbb{TM}) =  \mathbb{L}^+ \otimes \mathbb{L}^- \otimes \mathbb{S}^+ \otimes  \mathbb{S}^-, \]
\[ \Omega_3(\mathbb{M}) =  \mathbb{L}_+ \otimes \mathbb{L}_- \otimes \mathbb{S}_+ \otimes   \mathbb{S}_-, \]
\[ \Omega^4(\mathbb{TM}) =  (\mathbb{L}^+)^2 \otimes (\mathbb{L}^-)^2,\]
\[ \Omega_4(\mathbb{M}) =  \mathbb{L}^2_+ \otimes \mathbb{L}_-^2.\]
Traditionally, for a Penrose structure, one also requires that there be given an isomorphism of $\mathbb{L}^+$ and $\mathbb{L}^-$ and dually an isomorphism of $\mathbb{L}_+$ and $\mathbb{L}_-$.   We will not require this here. \\\\ The part $[G] = \mathbb{L}_+\otimes \mathbb{L}_-$ of $(\mathbb{T^*M})^2$ gives $\mathbb{M}$ a canonical conformal structure, whose inverse is the part $\mathbb{L}^+\otimes \mathbb{L}^-$ of the bundle $(\mathbb{TM})^2$.
\eject\noindent
To construct a null tetrad, we take a basis $\{\alpha, \beta\}$ of $\mathbb{S}_+$ and a basis $\{\gamma, \delta\}$ of $\mathbb{S}_-$.   Then the four co-vectors $l  = \theta(\alpha\otimes \gamma), m = \theta(\beta\otimes \gamma),  m' = \theta(\alpha\otimes \delta),  n = \theta(\beta\otimes \delta)$ form a null tetrad and $g = 2(ln - mm')$ represents the conformal structure.    A general co-vector $T$ is represented by $u \alpha\otimes \beta + x \beta\otimes \gamma + y \alpha\otimes \delta + v \beta\otimes \delta$.  The image of $T^2$ in $\mathbb{L}_+ \otimes \mathbb{L}_-$ is then:
\[ g^{-1}(T, T) = 2(uv - xy) (\alpha\wedge\beta)\otimes (\gamma\wedge \delta).\]
In particular for any $p$ and $q$ not both zero, the vectors $pl + qm = p\alpha\otimes\gamma + q\beta\otimes \gamma = (p\alpha + q\beta)\otimes \gamma$ and $pm' + qn =  p\alpha\otimes \delta+ q\beta \otimes \delta = (p\alpha+ q\beta) \otimes \delta$ span an isotropic plane.  So the isotropic planes of one type are parametrized by the elements of  $\mathbb{PS}_+$.  Similarly the co-vectors $sl + tm' = s\alpha\otimes \gamma + t\alpha\otimes \delta =  \alpha\otimes (s\gamma + t\delta)$ and $sm + tn =  s\beta\otimes\gamma + t\beta\otimes \delta = \beta\otimes (s\gamma + t\delta)$ span an isotropic plane of the other type.  So we see that the isotropic planes of the other type are parametrized by the elements of  $\mathbb{PS}_-$.     
Focussing on the first kind of isotropic plane, we have:
\[ (pl + qm)(pm' + qn) = p^2 lm' + pq(ln + mm) + q^2 mn\]
\[ = (p\alpha+ q\beta) \otimes (\gamma\wedge \delta).\]
So if we take $\Sigma^+$ to be spanned by the forms $lm', ln + mm', mn$, we see that the null cone bundle of the associated Weyl structure can be identified with the bundle $\mathbb{PS}_+$.  Then the null cone bundle of $(\mathbb{M}, \Sigma^-)$ can be identified with $\mathbb{PS}_-$.  Note that henceforth we usually identify $\mathbb{T^*M}$ with $\mathbb{S}^+ \otimes \mathbb{S}^-$, reserving the use of $\theta$ to convert to forms.
\eject\noindent
\section{The local Levi-Civita connection}
Given the null tetrad $\{l = \alpha\otimes \gamma, m = \beta\otimes \gamma, m' = \alpha\otimes \delta, n = \beta\otimes \delta\}$, the associated Levi-Civita connection, $d$ of $\mathbb{M}$ is given by the formulas:
\[ d\otimes l = A \otimes l + B \otimes m + B'\otimes  m', \]
\[ d\otimes m = D\otimes l + E\otimes m + B'\otimes n, \]
\[ d\otimes m' = D'\otimes l - E\otimes m' + B\otimes n, \] 
\[ d\otimes n = D'\otimes m + D\otimes m' - A\otimes n.\]
Then $d$ automatically preserves the metric $g = 2(ln - mm')$ and these formulas give the general such connection; the torsion-free condition that fixes the connection one-forms $A$, $B$, $B'$, $D$ and $E$, uniquely and gives the Levi-Civita conneciton is:
\[ dl = Al + Bm + B'm', \]
\[ dm = Dl +Em + B'n, \]
\[ dm' = D'l - Em' + Bn, \]
\[ dn = D'm + Dm' - An.\]
Put $2P = A - E$ and $2Q = A + E$, so $A = P +Q$ and $E = Q - P$.  Then the connection $d$ lifts uniquely to the spin bundles, such that $d\otimes (\alpha\wedge\beta) = 0$ and $d\otimes (\gamma \wedge \delta) = 0$ (so in particular, the connection on the line bundles $\mathbb{L}_\pm$ and $\mathbb{L}^\pm$ is flat).  Explicitly we have the defining formulas:
\[ d\otimes \alpha = P\otimes \alpha + B\otimes \beta, \]
\[ d\otimes \beta = D \otimes \alpha - P\otimes \beta, \]
\[ d\otimes \gamma =  Q\otimes \gamma + B'\otimes \delta, \]
\[ d\otimes \delta = D'\otimes \gamma -  Q \otimes \delta. \]
 
\eject\noindent
\section{Abstract indices}  
To organize spinor calculations, we use the abstract index formalism of Penrose.  So here elements (or local sections), $v$ of $\mathbb{S}^+$, $w$ of $\mathbb{S}^-$, $x$ of $\mathbb{S}_+$ and $y$ of $\mathbb{S}_-$ are represented by $v^{A'}$ (a primed spinor), $w^A$ (an un-primed spinor), $x_{A'}$ (a primed co-spinor) and $y_{A}$ (an un-primed co-spinor).   Idempotent symmetrization is represented by parentheses around the indices being symmetrized.   Idempotent skew symmetrization is represented by brackets around the indices being skew-symmetrized.  Vector indices for tensors of $\mathbb{M}$ are pairs $a = AA'$, $b = BB'$, etc.  The ordering of primed indices relative to unprimed indices is immaterial.  The natural skew mappings on the spin spaces are  written $\epsilon_{A'B'}$, $\epsilon_{AB}$, $\epsilon^{A'B'}$ and $\epsilon^{AB}$,  each a skew tensor, taking values $\mathbb{L}^+$, $\mathbb{L}^-$, $\mathbb{L}_+$ and $\mathbb{L}_-$, respectively,  with the relative normalization given by $\epsilon^{A'B'}\epsilon_{A'B'} = \epsilon^{AB}\epsilon_{AB} = 2$; also we have $\epsilon^{AB}\epsilon_{AC} = \delta_C^B$, $\epsilon^{AB}\epsilon_{CD} =  2\delta^A_{[C}\delta^B_{D]}$,  $\epsilon^{A'B'}\epsilon_{A'C'} = \delta_{C'}^{B'}$ and $\epsilon^{A'B'}\epsilon_{C'D'} =  2\delta^{A'}_{[C'}\delta^{B'}_{D']}$,  where $\delta_B^C$ and $\delta_{B'}^{C'}$ are the Kronecker delta spinors.   Indices are raised and lowered with these skew spinors, according to the rules:
\[ v_{A'} =  v^{B'} \epsilon_{B'A'}, \hspace{10pt} w_A = w^B \epsilon_{BA},  \hspace{10pt} x^{A'} =   \epsilon^{A'B'}x_{B'}, \hspace{10pt} y^A = \epsilon^{AB}y_B.\]  
The canonical one-form is $\theta^a = \theta^{AA'}$.  The conformal structure is then:
\[ 4[g] = \theta^{[A'[A}\otimes \theta^{B]B']} =  \epsilon^{A'B'}\epsilon^{AB} g, \hspace{10pt} g = \theta^a \otimes \theta_a = \theta^{AA'} \otimes \theta^{BB'}\epsilon_{AB} \epsilon_{A'B'}.\]
Any spinor may be decomposed into irreducible parts, each of which is totally symmetric and totally trace-free. In particular, we have the decomposition:
\[ \theta^a \theta^b =  \epsilon^{AB} \Sigma^{A'B'} + \epsilon^{A'B'} \Sigma^{AB}, \]
\[ \Sigma^{A'B'} = \frac{1}{2} \epsilon_{AB} \theta^a \theta^b = \Sigma^{B'A'}, \hspace{10pt} \Sigma^{AB} = \frac{1}{2} \epsilon_{A'B'} \theta^a \theta^b = \Sigma^{BA}. \]
Then $\Sigma^{A'B'}$ is called self-dual and takes values in $\mathbb{L}^-$, whereas  $\Sigma^{AB}$ is called anti-self-dual and takes values in $\mathbb{L}^+$.  For a two-form $\omega = \omega_{ab} \theta^a \theta^b$, where $\omega_{ab} = - \omega_{ba}$, we have its conformally invariant decomposition into self-dual and anti-self-dual parts:
\[ \omega = \omega_+ + \omega_-, \]
\[ \omega_+ = \omega_{A'B'} \Sigma^{A'B'}, \hspace{10pt} \omega_{A'B'} = \epsilon^{AB} \omega_{ab},  \]
\[ \omega_- = \omega_{AB} \Sigma^{AB}, \hspace{10pt} \omega_{AB} = \epsilon^{A'B'} \omega_{ab},   \]
\[ \omega_{ab}= \frac{1}{2} \epsilon_{AB}\omega_{A'B'} +  \frac{1}{2} \epsilon_{A'B'}\omega_{AB}.\]
\eject\noindent
\section{Spin connections}
The spin connection corresponding to the local Levi-Civita connection acts on the tensor algebra generated by the spin spaces $\mathbb{S}^\pm$ and their duals.  The connection respects duality; the torsion-free condition is the condition:
\[ d\theta^{a} = 0.\]
Further the curvature of the induced connections on the line-bundles $\mathbb{L}^\pm$ and $\mathbb{L}_\pm $ vanishes.  The curvature two-forms $R_{A'B'} = R_{B'A'} $ and $R_{AB} = R_{BA}$ are given by the formulas, valid for any spinors fields $v_{A'}$ and $v_A$:
\[ d^2 v_{A'} = R_{A'B'} v^{B'}, \hspace{10pt} d^2 v_A =  R_{AB} v^B.\]
The Riemann two-form, $R_{ab} = - R_{ba} = \theta^c \theta^d R_{cdab}$, which is defined by the relation $d^2 v_a = - R_{ab}v^b$,  valid for any co-vector field $v_a$,  is given by the formula:
\[ R_{ab} = R_{A'B'} \epsilon_{AB} + R_{AB} \epsilon_{A'B'}.\]
Here we may write:
\[ R_{A'B'} = \theta^c \theta^d R_{cdA'B'}, \hspace{10pt} R_{AB} = \theta^c \theta^d R_{cdAB}, \]
\[ R_{cdA'B'} = R_{[cd](A'B')}, \hspace{10pt} R_{cdAB} = R_{[cd](AB)}.\]
Then we have:
\[ R_{cdab} = R_{cdA'B'}\epsilon_{AB} + R_{cdAB}\epsilon_{A'B'}.\]
The first Bianchi identity is:
\[ 0 = d^2 \theta^a = - R^{ab} \theta_b = R_{B'}^{A'} \theta^{AB'} +  R_{B}^{A} \theta^{A'B}.\]
Using the first Bianchi identity, the Riemann form decomposes as:
\[ R^{ab} = C^{ab} + 2\theta^{[a} S^{b]} - \Lambda \theta^a \theta^b,\]
\[ \delta_b C^{ab} = 0, \hspace{10pt} \delta_a S^a = 0.\]
Here $\delta_b$ is the derivation of forms of degree minus one, such that $\delta_b \theta^a = \delta_b^a$.  Then  $C_{ab} = - C_{ba} = \theta^c\theta^d C_{cdab}$ is called the Weyl form and $C_{abc}^{\hspace{12pt}d}$ the Weyl tensor.  The form $S_a = \theta^b S_{ab}$ represents (up to sign) the trace-free part of the Ricci tensor.   The Ricci form is:
\[ R^a =  \frac{1}{2} \delta_bR^{ab} = -  S^{a}  +\frac{3}{2} \Lambda  \theta^a.\]
The Ricci scalar is $\delta_a R^a = 6\Lambda$.  
\eject\noindent We write out the various curvature forms, using the following formulas, which define the two-forms $\Sigma^{A'B'}$ and $\Sigma^{AB}$, the three-form $\Sigma_a$ and the four-form $\Sigma$:
\[ \theta^a \theta^b = \epsilon^{AB} \Sigma^{A'B'} + \epsilon^{A'B'} \Sigma^{AB}, \hspace{10pt} \Sigma^{A'B'} = \Sigma^{B'A'}, \hspace{10pt} \Sigma^{AB} = \Sigma^{BA}, \]
\[ \theta^a \Sigma^{B'C'} = \epsilon^{A'(B'}\Sigma^{C')A}, \hspace{10pt} \theta^a \Sigma^{BC} = - \epsilon^{A(B}\Sigma^{C)A'},\hspace{10pt} \theta^a \Sigma_b =  - \delta^a_b \Sigma.\] 
The Einstein three-form is:
\[ G^a = R_{B'}^{A'} \theta^{AB'} - R_{B}^{A} \theta^{A'B} = -2S^{ab}\Sigma_b - 3\Lambda \Sigma_a.\]
The second Bianchi identity is:
\[ dR_{ab} = 0, \hspace{10pt} dR_{A'B'} = 0, \hspace{10pt} d R_{AB} = 0.\]
The second Bianchi identity implies, in particular, the Einstein conservation law:
\[ 0 = dG^a.\]
Using the information of the first Bianchi identity, the decomposition of the curvature spinors can be written out as follows:
\[ R_{A'B'} = - \Sigma^{C'D'} C_{A'B'C'D'} - \Lambda \Sigma_{A'B'} - \Sigma^{AB} S_{ab},\]
\[ R_{AB} = - \Sigma^{CD} C_{ABCD} - \Lambda \Sigma_{AB} - \Sigma^{A'B'} S_{ab}.\]
Here the coefficients $C_{A'B'C'D'}$, called the self-dual Weyl spinor, $C_{ABCD}$, the anti-self-dual Weyl spinor and $S_{ab} = S_{ABA'B'}$, the trace-free Ricci tensor, are totally symmetric spinors.  For the tensor $S_{ab}$ this means that $S_{ab}$ is symmetric and trace-free.
Also we have for the Weyl form:
\[ C^{ab} = C^{AB} \epsilon^{A'B'} + C^{A'B'} \epsilon^{AB}, \]
\[ C_{AB}  = - \Sigma^{CD} C_{ABCD}, \hspace{10pt} C_{A'B'} =  - \Sigma^{C'D'} C_{A'B'C'D'}.\]
Finally we note the relations:
\[ \epsilon^{CD}R_{cdA'B'} = - C_{A'B'C'D'}  + \Lambda \epsilon_{A'(C'} \epsilon_{D')B'},\]
\[ \epsilon^{C'D'}R_{cdA'B'} = - C_{ABCD}  + \Lambda \epsilon_{A(C} \epsilon_{D)B}.\]
It is a basic property of the local Levi-Civita connections that the Weyl tensor $C_{abc}^{\hspace{12pt} d}$ depends only on the conformal structure not on the particular Levi-Civita connection used to obtain it.  It follows that the Weyl spinors are conformally invariant also.
\eject\noindent
\section{The lift to the co-spin bundle}
We lift the connection to the co-spin bundle $\mathbb{S}_+ $.  This gives a system of horizontal vector fields, denoted $\partial_a$.  Dually, we have a tautological indexed section $\pi_{A'}$ of $\mathbb{S}_+$, whose value at $(x, \pi_{A'})$, with $x \in \mathbb{M}$ is $\pi_{A'}$.  The section $\pi_{A'}$ takes values in the pull-back of $\mathbb{S}_+$ to itself.  More formally, the pull-back bundle is the collection of all triples $(x, \pi, \eta)$ with $(x, \pi) \in \mathbb{S}_+$ and $(x, \eta) \in \mathbb{S}_+$, with fiber map: $(x, \pi, \eta) \rightarrow (x, \pi)$.  The tautological section is then all triples of the form $(x, \pi, \pi)$, with $(x, \pi) \in \mathbb{S}_+$.  Then the connection gives a one-form $d\pi_{A'}$ with values in the same bundle.  A complete system of holomorphic one-forms for the spin bundle is then $\theta^{AA'}$ and $d\pi_{A'}$, with the  exterior derivatives:
\[ d\theta^{AA'} = 0, \hspace{10pt} d(d\pi_{A'}) = d^2 \pi_{A'} =  R_{A'B'} \pi^{B'}.\]
Relative to the spin basis $\alpha, \beta$ used above, we have:
\[ \pi_{A'} = p\alpha_{A'} + q \beta_{A'}, \]
\[ d\pi_{A'} = (dp)\alpha_{A'} + dq \beta_{A'} + p d\alpha_{A'} + qd\beta_{A'} \]
\[ = \left(dp +pP+ qD\right) \alpha_{A'} +  \left(dq + pB - qP\right) \beta_{A'}.\]
Here the variables $p$ and $q$ serve as fiber co-ordinates.    Then we have:
\[ \pi^{A'} d\pi_{A'} =  qdp- pdq - p^2B + 2pqP + q^2D.\]
\eject\noindent
\section{The spinor description of the Weyl connections}
Now consider the one-forms $\theta^{AA'} \pi_{A'}$ and $\pi^{A'} d\pi_{A'}$; these span a three-dimensional sub-bundle $\mathcal{T}'$ of the co-tangent bundle of $\mathbb{S}_+$; but they are homogeneous, so induce a three-dimensional sub-bundle of the sphere bundle $\mathbb{PS}_+$, also called $\mathcal{T}'$.   Recall the spinor bases $\{ \alpha_{A'}, \beta_{A'}\}$ and $\{\gamma_A, \delta_A\}$, used above.  Note that we have:
\[ \gamma_A\pi_{A'}\theta^{AA'}  =  p\alpha_{A'}\gamma_A\theta^{AA'} + q\beta_{A'}\gamma_A\theta^{AA'} = p \alpha\otimes \gamma + q\beta\otimes \gamma = pl + qm, \]
\[ \delta_A\pi_{A'}\theta^{AA'}  =  p\alpha_{A'}\delta_A\theta^{AA'}+ q\beta_{A'}\delta_A \theta^{AA'} = p\alpha\otimes \delta + q\beta\otimes \delta = pm' + qn.\]
This shows that $\mathcal{T}'$ gives a Weyl connection for $\mathbb{PS}_+$.  We compute the Weyl torsion of this connection, using the fact that $\theta^{AA'} = \pi^{A'} \phi^A$, mod $\theta^a \pi_{A'}$, for some one-form $\phi^A$.  Note that then we have:
\[ l = \alpha_{A'} \gamma_{A} \theta^{a} = \alpha_{A'} \pi^{A'} \gamma_A\theta^a = q\gamma_A\phi^A = -q\phi, \]
\[ m = \beta_{A'} \gamma_{A} \theta^{a} = \beta_{A'} \pi_{A'} \gamma_A\theta^a = -p\gamma_A\phi^A = p\phi, \]
\[ m' = \alpha_{A'} \delta_{A} \theta^{a}  =  \alpha_{A'} \pi^{A'} \delta_A\theta^a  = q\delta_A\phi^A = q\psi, \]
\[ n = \beta_{A'} \delta_{A} \theta^{a}  =  \beta_{A'} \pi_{A'} \delta_A\theta^a  = -p\delta_A\phi^A = -p\psi. \]
So we have $\phi^A \gamma_A = - \phi$ and $\phi^A\delta_A = \psi$, so we may write $\phi^A = - \psi \gamma^A - \phi \delta^A$.  Then we have $\phi_A\phi^A = - 2 \phi \psi = - 2\sigma$.  Now for the torsion, we compute $d(\theta^{AA'} \pi_{A'})  $ mod $\mathcal{T'}$.  This gives, since $d\theta^a = 0$:
\[ d(\theta^{AA'} \pi_{A'})  = - \theta^{AA'}  d\pi_{A'}  = - \phi^A (\pi^{A'} d\pi_{A'})   = 0, \mod \mathcal{T}'.\]
Thus the Weyl torsion vanishes.  Since we have shown in section six above that the torsion-free Weyl connection is unique, this shows that the present construction gives the torsion-free Weyl connection.  Next we compute the Weyl curvature of the Weyl connection.   Using the fact that $d\pi_{A'} = \pi_{A'} \lambda$, for some one-form $\lambda$, modulo $\mathcal{T'}$, it follows that  $(d\pi^{A'}) d\pi_{A'} = \pi^{A'} \pi_{A'} \lambda^2 = 0$, modulo $\mathcal{T}'$, so we have, working modulo $\mathcal{T'}$:
\[ d(\pi^{A'} d\pi_{A'}) =  (d\pi^{A'}) d\pi_{A'} + \pi^{A'} \theta^c\theta^d R_{cdA'B'} \pi^{B'}  =  \pi^{A'} \pi^{B'}\pi^{C'}\pi^{D'}\phi^C\phi^DR_{cdA'B'} \]
\[ = \frac{1}{2}\phi_A\phi^A \pi^{A'} \pi^{B'}\pi^{C'}\pi^{D'} \epsilon^{CD} R_{cdA'B'} = - \sigma \pi^{A'} \pi^{B'}\pi^{C'}\pi^{D'} \epsilon^{CD} R_{cdA'B'} \]
\[ = - \sigma \pi^{A'} \pi^{B'}\pi^{C'}\pi^{D'}(- C_{A'B'C'D'}  + \Lambda \epsilon_{A'(C'} \epsilon_{D')B'})  = \sigma \pi^{A'} \pi^{B'}\pi^{C'}\pi^{D'}C_{A'B'C'D'}.\] 
In the last section, we gave the expressed  the one-form $\pi^{A'} d\pi_{A'}$ as  $qdp- pdq - p^2B + 2pqP + q^2D$ in terms of the co-ordinates $p$ and $q$.  Comparing with the co-ordinate expression for the one-form $\gamma$, we see that necessarily we have the relation: $\gamma = \pi^{A'}d\pi_{A'} $ modulo $\Theta$.  Therefore we see that for the Weyl curvature, $W = d\gamma$ mod $\mathcal{T}'$, of the Weyl connection, we have the expression:
 \[  W = C_{A'B'C'D'} \pi^{A'} \pi^{B'} \pi^{C'} \pi^{D'}.\]
 As expected $W$ is homogeneous of degree four in the spinor $\pi_{A'}$.  Also this interprets the curvature $W$ of the Weyl connection as encoding the information of the self-dual Weyl spinor of the conformal structure.  In particular we have the lemmas:
 \begin{lemma} The torsion-free Weyl structure, on the primed projective spin bundle $\mathbb{PS}_+$, associated to a given conformal structure is a twistor structure if and only if the self-dual part of the Weyl curvature, $C_{A'B'C'D'}$ vanishes identically.
 \end{lemma}
 \begin{lemma} The torsion-free Weyl structure, on the primed projective spin bundle $\mathbb{PS}_+$, associated to a given conformal structure, which has non-vanishing self-dual Weyl curvature, $C_{A'B'C'D'}$, is naturally a $(2, 3, 5)$-system, on the complement of the zero set of the homogeneous function on the projective spin bundle $\mathbb{PS}_+$, $W = C_{A'B'C'D'} \pi^{A'}\pi^{B'}\pi^{C'} \pi^{D'}$.
 \end{lemma}
 \eject\noindent
\section{The $\mathbb{G}_2$-conformal structure}
After Cartan, Nurowski, \v{C}ap and Sagerschnig, we know that given any $(2, 3, 5)$ structure on five manifold, $\mathcal{S}$, there is canonically associated a conformal structure on $\mathcal{S}$, which has holonomy in $\mathbb{G}_2$.  It is possible to  determine this conformal structure for the case of the torsion-free Weyl structures of conformal structures, whose self-dual Weyl curvature is not identically zero. To express the result, we use a local Levi-Civita connection, as described above.    Introduce the differential operator on the primed spin-bundle $\mathbb{S}_+$, $D_A = \pi^{A'} \partial_a$, where $\partial_a$ are the horizontal vector-fields of the connection (so $\theta^a(\partial_b) = \delta^a_b$ and $(d\pi_{A'})(\partial_a) =0$).  Then define:
\[ \psi = C_{A'B'C'D'} \pi^{A'}\pi^{B'}\pi^{C'} \pi^{D'}, \hspace{10pt} \psi_A = \psi^{-1} D_A \psi, \hspace{10pt} \psi_{AB} = \psi^{-1} D_A D_B \psi.\]
Note that $\psi = W$, where $W$ is the Weyl curvature of the Weyl connection and we work only at points where $\psi \ne 0$; it follows from the definition of $\psi$ and formulas for the curvature that $\psi_{AB}$ is automatically symmetric.  Also $\psi, \psi_A$ and $\psi_{AB}$ are homogeneous of degrees four, one and two, respectively, in the variable $\pi_{A'}$.  Recall that the pieces of the primed curvature are given by the formulas:
\[ d^2 \pi_{A'} = R_{A'B'} \pi^{B'}, \hspace{10pt} R_{A'B'} = - \Sigma^{C'D'} C_{A'B'C'D'} - \Lambda \Sigma_{A'B'} - \Sigma^{AB} S_{ab}.\]
Define the one form, $\theta^A$ and the spinor $\tau_{AB} = \tau_{BA}$ by the formulas:
\[ \theta^A = \theta^a\pi_{A'}, \hspace{10pt} \tau_{AB} = \frac{1}{40\psi} (-16\pi^{A'}\pi^{B'} S_{ab}  + 5\psi_A \psi_B - 4\psi_{AB}).\]
Note that $\theta^A$ and $\tau_{AB}$ are homogeneous of degrees one and minus two in the variable $\pi_{A'}$, respectively.   The main result of this work then is:
\begin{theorem}  The canonical conformal structure $\mathcal{G}$ on the sphere bundle of the torsion-free Weyl structure of a conformal structure in four-dimensions may be given by the formula:
\[ \mathcal{G} = \frac{1}{12\psi}  (4\pi^{A'} d\pi_{A'} - \psi_A\theta^A)^2 + \theta^a\theta_a + 2\theta^A \theta^B \tau_{AB}.\]
\end{theorem}
\noindent The proof of the theorem is by direct calculation, following \v{C}ap and Sagerschnig, and will be given in sections fifteen to twenty below.  The result can be stringently tested, in examples, by calculating this conformal structure, by first converting into the canonical form of Cartan and then using Nurowski's formula.  Here we confine ourselves to two examples, one, the case of the Kapadia family of plane wave metrics, which contains the flat $\mathbb{G}_2$ geometry as a special case, and the other the famous Schwarzschild metric.
\eject\noindent
\section{Summary of key spin-bundle formulas}
The spin-connection $\partial_a$ of the space-time is torsion free and kills $g_{ab}$,  $\epsilon_{AB}$ and $\epsilon_{A'B'}$.  The spinor curvature $R_{abC'D'}$ obeys the relations, for any spinor fields $v_{A'}$ and $v_A$ on the space-time:
\[ [\partial_a, \partial_b] v_{C'} = R_{abC'D'}v^{D'}, \hspace{10pt} [\partial_a, \partial_b] v_{C} = R_{abCD}v^{D},\]
\[ R_{abC'D'}  =   - \epsilon_{AB}C_{A'B'C'D'} -  \epsilon_{A'B'}S_{ABC'D'} +  \epsilon_{AB}  \Lambda\epsilon_{C'(A'} \epsilon_{B')D'}.\]
\[ R_{abCD}  =   - \epsilon_{A'B'}C_{ABCD} -  \epsilon_{AB}S_{A'B'CD} +  \epsilon_{A'B'} \Lambda \epsilon_{C(A} \epsilon_{B)D}.\]
Lifting to the spin bundle we get the following commutators of derivatives:
\[ [\partial^{A'}, \partial^{B'}] = 0, \hspace{10pt} [\partial_a, \partial^{B'}] = 0,  \]
\[ [\partial_a, \partial_b]   = - R_{abC'D'} \pi^{C'}\partial^{D'} =  \epsilon_{AB}  \Lambda \pi_{(A'} \partial_{B')} + \epsilon_{AB}C_{A'B'C'D'}\pi^{C'} \partial^{D'} +  \epsilon_{A'B'}S_{ABC'D'}\pi^{C'} \partial^{D'}.\]
Here we act on functions on the spin bundle holomorphic in the spinor $\pi_{A'}$.  The spinors  $C_{A'B'C'D'}$ and $C_{ABCD}$ are totally symmetric.  For a real space-time, the function $\Lambda$ and the trace-free symmetric tensor $S_{ab} = S_{ABA'B'} = S_{(AB)(A'B')}$ are real and $C_{ABCD}$ is the complex conjugate of $C_{A'B'C'D'}$.  The vertical vector field $\partial^{A'}$ annihilates functions and tensors pulled back to the spin bundle from the space-time.  Also we have the derivatives:
\[ \partial_a \pi_{B'} = 0, \hspace{10pt} \partial^{A'}\pi_{B'} = \delta^{A'}_{B'}. \] Next put $D_A = \pi^{A'} \partial_a$, so $D_A$ spans the (2, 3, 5)-structure.  Also define  the spinor $\psi_{A'} = C_{A'B'C'D'}\pi^{B'}\pi^{C'}\pi^{D'}$ and the scalar $\psi =  \psi_{A'} \pi^{A'} = C_{A'B'C'D'}\pi^{A'}\pi^{B'}\pi^{C'}\pi^{D'}$.   When $\psi \ne 0$, put $\psi_A = \psi^{-1} D_A \psi$ and $\psi_{AB} = \psi^{-1} D_AD_B \psi = \psi_{BA}$.  We have:, acting on (holomorphic)  functions on the spin-bundle:
\[ [D_A, D_B]  = \epsilon_{AB} \psi_{A'} \partial^{A'}, \hspace{10pt} D_CD^C = \psi_{A'}\partial^{A'}.\]
The Lie bracket of  vector fields  $V = v^a \partial_a + v_{A'} \partial^{A'}$ and $W = w^a \partial_a + w_{A'} \partial^{A'}$ is:
\[ [V, W] = (V(w^a) - W(v^a))\partial_a  + (V(w_{B'}) - W( v_{B'}) - v^c w^d\pi^{A'}R_{cdA'B'}) \partial^{B'}.\]
In the particular case that $v^a = v^A \pi^{A'}$ and $w^a = w^A \pi^{A'}$, so $V = v^AD_A + v_{A'} \partial^{A'}$ and $W = w^AD_A +w_{A'} \partial^{A'}$, the commutator simplifies to:
\[ [V, W] = (V(w^A) - W(v^A))D_A  + (V(w_{B'}) - W( v_{B'}) + v_Cw^C\psi_{B'}) \partial^{B'}.\]
Put $\theta = \pi^{C'} d\pi_{C'}$.  Then the covariant exterior derivative, $d$, obeys, in particular the relation: 
\[ d\theta = d(\pi^{A'}d\pi_{A'}) \]\[= (d\pi^{A'}) d\pi_{A'}  - \Sigma^{A'B'}(\Lambda \pi_{A'} \pi_{B'} + C_{A'B'C'D'}\pi^{C'}\pi^{D'}) - \Sigma^{AB}S_{ab}\pi^{A'}\pi^{B'}.\]
\eject\noindent
\section{The Reeb vector field and the contact form}
 We work systematically through the paper of \v{C}ap and Sagerschnig, either using their notation, or simple variants of their notation \cite{cap}.  We first need the Reeb vector field and the contact form.
Let $V = v^a \partial_a + v_{A'}\partial^{A'}$ be a homogeneous vector field on the spin bundle (with $v^a$ homogeneous of degree zero and $v_{A'} $ homogeneous of degree one, where $(v^a, v_{A'})$ is equivalent to $(v^a, v_{A'} + s\pi_{A'})$, for any function $s$ of degree zero).
\begin{itemize} \item $V$ belongs to $\mathbb{T}^{-1}$ iff  $v^a \pi_{A'} = 0$ and $v_{A'} = 0$ iff $v_{A'} = 0$ and $v^a = v^A \pi^{A'}$ for some unique $v^A$, iff $V = v^AD_A$.
 \item  $V$ belongs to $\mathbb{T}^{-2}$ iff $v^a \pi_{A'} = 0$ iff $v^a = v^A \pi^{A'}$ for some unique $v^A$ iff $V = v^AD_A + v_{A'} \partial^{A'}$.
 \item The map $q_{-2}$ may be construed as mapping the pair $(v^A\pi^{A'},  v_{A'})$ to $v_{A'}$ mod $\pi_{A'}$ or just to $v = v_{A'}\pi^{A'}$. So here $q_{-2}$ may be identified with contraction with the one-form on the spin-bundle, $\theta = \pi^{C'}d\pi_{C'}$.
 \item The map $q_{-3}$ may be construed as mapping the pair $(v^a, v_{A'})$ to $v^{a}\pi_{A'}$.  In particular $q_{-3}$ annihilates all vertical vector fields $v_{A'} \partial^{A'}$ and annihilates $\mathbb{T}^{-1}$ and $\mathbb{T}^{-2}$.  Since the image of $q_{-3}$ carries an index, we rewrite $q_{-3}$ as $q_{-3}^A$.  So here $q^A_{-3}$ may be identified with contraction with the one-form on the spin-bundle, $\theta^A = \pi_{A'}\theta^a$.
 \end{itemize}
We consider the image under $\theta^A$ of the iterated bracket $[V, [W, X]]$, where $V$ and $W$ belong to $\mathbb{T}^{-1}$ and $X$ belongs to $\mathbb{T}^{-2}$, but not to $\mathbb{T}^{-1}$, so we may write:
 \[ V = v^a \partial_a,  \hspace{10pt} v^a \pi_{A'} = 0, \hspace{10pt} v^a = v^A \pi^{A'},  \] 
 \[ W = w^a \partial_a,  \hspace{10pt}w^a \pi_{A'} = 0,  \hspace{10pt} w^a = w^A \pi^{A'},   \]
 \[ X = x^a \partial_a + x_{A'} \partial^{A'}, \hspace{10pt} x^a \pi_{A'} = 0,  \hspace{10pt} x^a = \xi^A \pi^{A'},  \hspace{10pt}  x = x_{A'}\pi^{A'} \ne 0.\] 
 Note that $x = x_{A'} \pi^{A'} $ has degree two.  We need to calculate:
 \[ \theta^C([V, [W, X]]) = \theta^C([ v^a \partial_a, [ w^b \partial_b, x^c \partial_c + x_{C'}\partial^{C'}]]).\]
We first consider the terms involving one or more derivatives of $x^a$ or $x_{A'}$: 
 \[ \theta^C([V, W(x^a) \partial_a + W(x_{A'})\partial^{A'} - X(w^b) \partial_b - w^a x^b R_{abC'D'}\pi^{C'}\partial^{D'} ])\]
  \[ = (VW(x^c)  - W(x^a)\partial_a v^c -  W(x_{A'})\partial^{A'} v^c   - V(X(w^c)) ) \pi_{C'}\]
    \[ =   v^cW(x_{C'})    - V(X(w^c)  \pi_{C'})  =   v^cW(x_{C'})    + w^c V(x_{C'})\]
      \[ = v^C \pi^{C'}W(x_{C'})    + w^C \pi^{C'} V(x_{C'})  = v^C W(x)    + w^C  V(x).\]
 \eject\noindent
The terms not involving the derivatives of either $x^a$ or  $x_{A'}$ are then the contraction $x_{C'}Z^{c} + x^d Z_d^{C}$, where firstly $Z^{c}$ is given as :
    \[  Z^{c}=  \theta^C([ v^a \partial_a, [ w^b \partial_b, \partial^{C'}]])  = -  \theta^C([ v^a \partial_a, (\partial^{C'}w^b) \partial_b])\]
\[ =\theta^C(( (\partial^{C'}w^a) ( \partial_a v^b) - (v^a \partial_a(\partial^{C'}w^b)) )\partial_b)\]
\[ = (\partial^{C'}w^a)( \partial_a v^{B'C})\pi_{B'} - v^a \partial_a(\partial^{C'}w^{B'C})\pi_{B'} \]
\[ =   V (w^c) = \pi^{C'}V(w^C).  \]
Secondly $Z_d^{C}$ is given by the formula:
\[ Z_d^C =  \theta^C([ v^a \partial_a, [ w^b \partial_b, \partial_d]])\]
\[ =  \theta^C([ v^a \partial_a, - (\partial_d w^b) \partial_b - w^b R_{bdA'C'}\pi^{A'} \partial^{C'}])\]
\[ = \theta^C(- v^a (\partial_a\partial_d w^e)  +  (\partial_d w^b) (\partial_b v^e )+ w^b R_{bdA'C'} \pi^{A'} (\partial^{C'} v^e))\partial_e\]
\[ = - v^a (\partial_a\partial_d w^c) \pi_{C'} +   (\partial_d w^b) (\partial_b v^c )\pi_{C'}+  w^b R_{bdA'E'} \pi^{A'} (\partial^{E'} v^c)\pi_{C'}\]
\[ =  - w^bv^c R_{bdA'C'} \pi^{A'} \]\[ =   \pi^{A'} \pi^{B'}\pi^{C'} w^B v^C \epsilon_{BD} C_{B'D'A'C'} + w^B v^C \pi^{B'} \pi^{C'}\pi^{A'}\epsilon_{B'D'}   S_{BDA'C'}\]
\[ = v^C (w_D \psi_{D'} + \pi_{D'} w^B S_{DBA'B'}\pi^{A'}\pi^{B'}).\]
So, altogether, we have (since $x^d\pi_{D'} = 0$):
 \[ \theta^C([V, [W, X]]) =   v^C W(x)    + w^C  V(x) + x_{C'}\pi^{C'}V(w^C) + v^C w_D \psi_{D'}\xi^D\pi^{D'}\]
 \[ = x V (w^C) + v^C w^B(D_B(x)  - \xi_B \psi)   + w^C  v^BD_B(x) \]
 \[ = x (V(w^C) + v^A w^B\Gamma_{AB}^{\hspace{13pt}C}), \]
 \[ \Gamma_{AB}^{\hspace{13pt} C} = \delta_A^C (x^{-1} D_B x - x^{-1} \xi_B\psi) + \delta_B^C  x^{-1}D_A x.  \]
 Put $x_A  = x^{-1} D_A x$ and $x_{AB} = x^{-1} D_{(A}D_{B)} x$, so now we have:
 \[  \Gamma_{AB}^{\hspace{13pt} C} = \delta_A^C (x_B- x^{-1} \xi_B\psi) + \delta_B^C  x_A,  \]
 \[ \Gamma_{AB}^{\hspace{13pt} B} = 3x_A   - x^{-1} \xi_A \psi. \]
 Now let $\partial_V(W)= \mu^A\pi^{A'}\partial_a$.\\
 Then we have:
 \[ \{ \partial_V(W), \phi\}^A  = \theta^A([\mu^B\pi^{B'}\partial_b, X])\]
 \[ = - (x_{C'} \partial^{C'}(\mu^A\pi^{A'})) \pi_{A'}\]
 \[ = \mu^A x = \theta^A([V, [W, X]]). \]
 So we have:
 \[ \mu^A = \partial_V(W)^A = V(w^A) + v^B w^C\Gamma_{BC}^{\hspace{13pt}A}. \]
 \eject\noindent
We extend to $\textrm{gr}_{-2}$ by the formula, valid for $W_1 = w_1^A \pi^{A'} \partial_a$ and $W_2 = w_2^A \pi^{A'} \partial_a$:
\[ \partial_V(  (w_1)_A w_2^A \psi) = \textrm{gr}_{-2} ([(\partial_V W_1), W_2] -  [(\partial_V W_2), W_1])  = ((\partial_V W_1)_Cw_2^C - (\partial_V W_2)_Cw_1^C)\psi \]
  \[ = \psi(  V((w_1)_Aw_2^A) - v^A \Gamma_{AB}^{\hspace{13pt}C} w_1^B (w_2)_C + v^A \Gamma_{AB}^{\hspace{13pt}C} w_2^B (w_1)_C)\]
   \[ = \psi( V((w_1)_Aw_2^A) - v^A \Gamma_{AB}^{\hspace{13pt}B} w_1^E (w_2)_E)\]
   \[ = ( V((w_1)_Aw_2^A) + x^{-1}( 3V( x)  - v^B\xi_B \psi)  (w_1)_A w_2^A)\psi.\]  
   Putting $y = (w_1)_A w_2^A \psi$, we get:
   \[ y^{-1} \partial_V y = y^{-1}\psi V(y\psi^{-1}) + x^{-1}( 3V( x)  - v^B\xi_B \psi) \]\[= v^B(y^{-1}D_B y + 3x^{-1}D_B x - x^{-1}\xi_B \psi - \psi^{-1} D_B \psi).\]  
 In particular we need that $x$ be covariantly constant, so we need:
   \[  0 = x^{-1}\partial_V(x)  = v^A(4x^{-1}D_A x -  \psi^{-1}D_A\psi -x^{-1} \xi_A \psi).\]
 Since we need this to be true for all $V$ so for all $v^A$,  we get that given $x$, $\xi^A$ is determined by the formula:  
 \[ \xi^A = x\psi^{-1}(4x^{-1}D^A x -  \psi^{-1}D^A\psi) = x\psi^{-1}(4x^A - \psi^A).\] 
Back substituting, we get:
\[ \Gamma_{AB}^{\hspace{13pt}C} =   \delta_A^C (\psi_B - 3x_B) + \delta_B^C  x_A , \hspace{10pt}\Gamma_{AB}^{\hspace{13pt}B} =  \psi_A - x_A. \]
So now let $X = x^a\partial_a + x_{A'}\partial^{A'}$ be given, with $x_{A'} \pi^{A'} = x$, $x^a = \pi^{A'} \xi^A$ and $\xi^A = x\psi^{-1}(4x^A  -  \psi^A)$.   This is the so-called Reeb vector field.  Let the associated contact form be $\gamma = x^{-1} \pi^{C'}d\pi_{C'} + \gamma_A \pi_{A'} \theta^a$.  Note that $\iota_X \gamma = 1$, as required.  We fix $\gamma_A$, which has degree minus one, by the requirement that:
\[ 0 = \iota_X \iota_{D_A} d\gamma\]\[ = \iota_X  \iota_{D_A}  ( \pi^{C'}d\pi_{C'}\theta^a(x^{-2} \partial_a x + \partial_{A'} \gamma_A)  + \theta^a \theta^b (\partial_a \gamma_B)\pi_{B'} + (2x)^{-1}\theta^a \theta^b R_{ab C'D'}\pi^{C'} \pi^{D'})\] 
\[ =  - x^{-1} D_A x -  x\gamma_A + x^{-1} \xi_A \psi.\] 
So we get:
\[ \gamma_A = x^{-1}(- x^{-1} D_A x + x^{-1}\psi \xi_A) = x^{-1}(3x_A -   \psi_A).\]
So the contact form dual to the Reeb vector field is:
\[ \gamma = x^{-1}( \pi^{C'}d\pi_{C'} + (3x_A -  \psi_A) \pi_{A'} \theta^a).\]
\eject\noindent
\section{The decomposition of the tangent bundle}
Let $Z = z^a \partial_a + z_{A'} \partial^{A'} $ be a vector field.  We find $Z_1 \in \mathbb{T}^{-1}$ such that $Z - [X, Z_1] \in \mathbb{T}^{-2}$; equivalently $\theta^A(Z - [X, Z_1] ) = 0$.  We have:
\[ Z_1 = y^a \partial_a, \hspace{10pt} y^a \pi_{A'} = 0, \hspace{10pt} y^a = y^A \pi^{A'}, \]
\[ \theta^A(Z - [X, Z_1]) = z^a\pi_{A'} + \theta^A([y^b \partial_b,   x^c \partial_c + x_{C'} \partial^{C'})\]
\[ = z^a\pi_{A'} + (y^b \partial_b  x^a - x^b \partial_b y^a -  x_{C'} \partial^{C'} y^a )\pi_{A'} \]
\[ = z^a \pi_{A'} + x_{A'} y^a  = z^a \pi_{A'} +  xy^A.\]
So $y^A = - x^{-1} z^a \pi_{A'}$.  Then we have:
\[ Z - [X, Z_1] = z^a \partial_a + z_{A'} \partial^{A'} - [x^c\partial_c + x_{C'} \partial^{C'}, y^a \partial_a] \]
\[ = (z^a - x^c\partial_c y^a + y^c \partial_c x^a- x_{C'}\partial^{C'} y^a) \partial_a  + (z_{D'} + y^A D_A x_{D'}  - \xi_Cy^C \psi_{D'}) \partial^{D'}\]
Now we have, by construction that $z^A$ exists, so that:
\[ z^a - x^c\partial_c y^a + y^c \partial_c x^a- x_{C'}\partial^{C'} y^a = z^{A} \pi^{A'}.\]
Explicitly, we have:
\[ z^A = x^{-1}z^{AB'}x_{B'} - \xi^CD_C y^A + y^CD_C \xi^A - x_{C'}\partial^{C'} y^A. \]
Then we get:
\[  Z - [X, Z_1] - kX = (z^A- k\xi^A) D_A  + (- kx_{D'} + z_{D'} + y^A D_A x_{D'}  - \xi_Cy^C \psi_{D'}) \partial^{D'}\]
Choose $k$ to eliminate the last term: 
\[ k  =  x^{-1}(z_{D'}\pi^{D'} + y^A D_A x  - \xi_Cy^C \psi) =  x^{-1}z_{D'}\pi^{D'} + y^A (\psi_A - 3x_A).\]
Summarizing we have the decomposition:
\[ Z = [X, Z_1] + kX + Z_2, \]
\[ Z_1 = y^AD_A,  \hspace{10pt} Z_2 = u^AD_A, \]
\[  y^A = - x^{-1} z^a \pi_{A'}, \]
\[ u^A = x^{-1}z^{a}x_{A'} -  X(y^A) + y^BD_B \xi^A  - k\xi^A, \]
\[  k =   x^{-1}z_{D'}\pi^{D'} + y^A (\psi_A - 3x_A).\]
\eject\noindent
\section{The map $\Psi$}
We next calculate the map $\Psi$.
Acting on $T = t^a \partial_a = t^A D_A$, where $t^a = t^A \pi^{A'}$, $\Psi(T)$ is given by the formula:
\[ 2\{\Psi(T), q_{-3}(X)\}^A =  \theta^A([X, [T, X]])\]
\[ = \theta^A([X, [t^a\partial_a, x^b\partial_b + x_{B'} \partial^{B'}]])\]
\[ =  \theta^A([x^c \partial_c + x_{C'} \partial^{C'}, (T(x^b) - X(t^b))\partial_b + (T(x_{D'}) + t_F \xi^F\psi_{D'})\partial^{D'}]])\]
\[ =  x^c (\partial_c(T(x^a)\pi_{A'} - X(t^a)\pi_{A'})) +   (x_{C'} \partial^{C'} (T(x^a) - X(t^a)))\pi_{A'}  \]\[ -  (T(x_{D'}) + t_F \xi^F\psi_{D'} )(\partial^{D'}x^a)\pi_{A'}  - (T(x^b) - X(t^b))(\partial_b x^a)\pi_{A'}\]
\[ =  x^c \partial_c(t^ax_{A'}) +   (x_{C'} \partial^{C'} (T(x^a)\pi_{A'} - X(t^a)\pi_{A'})) 
-   x_{A'} (T(x^a) - X(t^a))\]
\[ + \xi^A\pi^{A'} (T(x_{A'}) + t_F x^F \psi_{A'} )\]
\[ =   X(x t^A)  -   x(T(\xi^A) - X(t^A)) + \xi^A(T(x) + t_F \xi^F \psi ) \]
\[ =  2x X( t^A) + t^A X(x)  -   xT(\xi^A) + \xi^A(T(x) + t_F \xi^F \psi) \]
\[ =2 \{\Psi(T)^a \partial_a, x_{E'} \partial^{E'}\}^A  =  - 2x_{F'} \partial^{F'} ( \Psi(T)^A \pi^{A'} )\pi_{A'}\]
\[ =  2x \Psi(T)^A.\]
So we get:
\[ \Psi(T)^A = X(t^A) +  P_B^{\hspace{5pt} A} t^B, \]
\[ P_B^{\hspace{5pt} A}  = \frac{1}{2} ( \delta_B^A x^{-1}X(x)  -   D_B(\xi^A) +  \xi^A(x_B - x^{-1}\psi \xi_B )) .\]
For the metric we need the symmetric part of $P_{AB}$. We have:
\[ 2P_{(AB)} =  -   D_{(A}\xi_{B)}+  x^{-1} \xi_{(A}D_{B)}x - x^{-1}\psi \xi_A\xi_B\]
\[ =  -   D_{(A}(4\psi^{-1}D_{B)} x -  x\psi^{-2}D_{B)}\psi)+  x^{-1} (4\psi^{-1}D_{(A} x -  x\psi^{-2}D_{(A}\psi)D_{B)}x - x^{-1}\psi \xi_A\xi_B\]
\[ =  5\psi^{-2} (D_{(A}x)D_{B)} \psi -   4\psi^{-1} D_{(A}D_{B)} x  + x\psi^{-2}D_{(A}D_{B)}\psi - 2x\psi^{-3}(D_{(A}\psi)D_{B)}\psi \]
\[ +  4x^{-1}\psi^{-1}(D_{(A} x) D_{B)}x -  \psi^{-2}(D_{(A}x )D_{B)}\psi - x\psi^{-1} (4x^{-1}D_A x -  \psi^{-1}D_A\psi)(4x^{-1}D_B x -  \psi^{-1}D_B\psi)\]
\[ =   x\psi^{-2}D_{(A}D_{B)}\psi-   4\psi^{-1} D_{(A}D_{B)} x   \] \[ +   12\psi^{-2} (D_{(A}x)D_{B)} \psi  - 3x\psi^{-3}(D_{A}\psi)D_{B}\psi  - 12x^{-1}\psi^{-1}(D_{A} x) D_{B}x. \]
So we get:
\[ P_{(AB)} =   \frac{x}{2\psi}(\psi^{-1}D_{(A}D_{B)}\psi-   4x^{-1} D_{(A}D_{B)} x   - 3(\psi^{-1}D_{A}\psi - 2x^{-1} D_A x)(\psi^{-1}D_{B}\psi - 2x^{-1}D_B x)) \]
\[ = \frac{x}{2\psi}(\psi_{AB}-   4x_{AB}   - 3(\psi_{A} - 2x_A)(\psi_{B} - 2x_B)). \]
\eject\noindent
\section{The map $\Phi$}
Next we calculate the map $\Phi$.
Let $V = v^a \partial_a$ and $W = w^a\partial_a$ in $\mathbb{T}^{-1}$ be given, with $v^a = v^A \pi^{A'}$ and $w^a = w^A \pi^{A'}$.  Then we have:
\[ \gamma([V, W]) = \gamma([v^a \partial_a, w^b \partial_b]) \]
\[ =  \gamma((V(w^B) - W(v^B)) D_B + v_Cw^C \psi_{B'}\partial^{B'}) \]
\[ = x^{-1} v_Cw^C \psi.\]
Also we have:
\[ Z = [V, W] - \gamma([V, W]) X \]
\[ = (V(w^A) - W(v^A)) D_A + v_Cw^C \psi_{B'}\partial^{B'} - x^{-1} v_Cw^C \psi (x_{A'}\partial^{A'} + \xi^AD_A) =  Z^A D_A, \]\[ \hspace{10pt} Z^A =  V(w^A) - W(v^A) + v_C w^CU^A,\hspace{10pt}  U^A =  - x^{-1} \psi \xi^A.\]
Now we have, for $T = t^a\partial_a$, where $t^a = t^A \pi^{A'}$:
\[ \partial_V\partial_W T - \partial_W\partial_V T - \partial_Z T\]
\[ =  \partial_V(W(t^A) + w^B t^C\Gamma_{BC}^{\hspace{13pt}A}) -  \partial_W(V(t^A) + v^B t^C\Gamma_{BC}^{\hspace{13pt}A}) - Z^BD_B t^A -  Z^Bt^C \Gamma_{BC}^{\hspace{13pt}A}\]
\[ =  (VW- WV)(t^A) + V(w^B t^C\Gamma_{BC}^{\hspace{13pt}A})  -  W(v^B t^C\Gamma_{BC}^{\hspace{13pt}A})\]\[  + v^B \Gamma_{BC}^{\hspace{13pt}A}(W(t^C) + w^D t^E\Gamma_{DE}^{\hspace{13pt}C})- w^B \Gamma_{BC}^{\hspace{13pt}A}(V(t^C) + v^D t^E\Gamma_{DE}^{\hspace{13pt}C}) -  Z^BD_B t^A -  Z^Bt^C \Gamma_{BC}^{\hspace{13pt}A}\]
\[ =  [v^b\partial_b,  w^c\partial_c](t^A)  - (V(w^B) - W(v^B))D_B t^A \]\[ +  (w^B t^C v^D -v^Bt^Cw^D)(D_D( \Gamma_{BC}^{\hspace{13pt}A}) -  \Gamma_{BE}^{\hspace{13pt}A}\Gamma_{DC}^{\hspace{13pt}E})    + x^{-1}\psi v_C w^C\xi^B(D_B t^A  +     t^C \Gamma_{BC}^{\hspace{13pt}A})\]
\[ =  v^Bw^C \pi^{B'} \pi^{C'} [\partial_b,  \partial_c](t^A) \]\[+ v_Ew^E (- t^C D^B \Gamma_{BC}^{\hspace{13pt}A}  +  t^E  \Gamma_{BC}^{\hspace{13pt}A}\Gamma_{\hspace{5pt}E}^{B\hspace{5pt}C}   +x^{-1} \psi \xi^BD_B t^A + x^{-1}\psi  \xi^Bt^C \Gamma_{BC}^{\hspace{13pt}A})\]
\[ =  v_F w^F( - \pi^{B'} \pi^{C'}S_{B'C'D}^{\hspace{26pt} A} t^D + \psi_{D'}\partial^{D'} t^A - t^C D^B \Gamma_{BC}^{\hspace{13pt}A})\]\[ + v_F w^F( t^E  \Gamma_{BC}^{\hspace{13pt}A}\Gamma_{\hspace{5pt}E}^{B\hspace{5pt}C}   +x^{-1} \psi \xi^BD_B t^A + x^{-1} \psi \xi^Bt^C \Gamma_{BC}^{\hspace{13pt}A}).\]
So we get:
\[ \Phi(T) =   x\psi^{-1}( -  \pi^{B'} \pi^{C'}S_{B'C'D}^{\hspace{26pt} A} t^D + \psi_{D'}\partial^{D'} t^A - t^C D^B \Gamma_{BC}^{\hspace{13pt}A} + t^E  \Gamma_{BC}^{\hspace{13pt}A}\Gamma_{\hspace{5pt}E}^{B\hspace{5pt}C}  )  +  \xi^BD_B t^A +  \xi^Bt^C \Gamma_{BC}^{\hspace{13pt}A}\]
\[ = X(t^A) +  Q_B^{\hspace{5pt} A} t^B, \]
\[ Q_B^{\hspace{5pt} A}  = \psi^{-1} x^{C'} \psi_{C'}\delta^A_B + \xi^C \Gamma_{CB}^{\hspace{13pt}A} +  x\psi^{-1}(  - \pi^{B'} \pi^{C'}S_{B'C'B}^{\hspace{26pt} A}   -  D^C \Gamma_{CB}^{\hspace{13pt}A} +   \Gamma_{CE}^{\hspace{13pt}A}\Gamma_{\hspace{5pt}B}^{C\hspace{5pt}E}  ).\]
\eject\noindent
For the metric, we need the symmetric part of $Q_{AB}$.  We have:
\[ \psi x^{-1} Q_{(AB)}  +  \pi^{A'} \pi^{B'}S_{ab}  = x^{-1} \psi \xi^C \Gamma_{C(BA)}   -  D^C \Gamma_{C(BA)} +   \Gamma_{CE(A}\Gamma_{\hspace{5pt}B)}^{C\hspace{5pt}E}.\]
\[ =  (4x_{(A}  -  \psi_{(A} )  (- 3x_{B)} + \psi_{B)})   -  D_{(A} (- 3x^{-1} D_{B)} x + \psi^{-1} D_{B)} \psi) \]
\[ +   (- 3x_E  + \psi_E) \epsilon_{C(A} \Gamma_{\hspace{5pt}B)}^{C\hspace{5pt}E}  +    x_C\epsilon_{E(A}\Gamma_{\hspace{5pt}B)}^{C\hspace{5pt}E}\]
\[ =   3x_{AB} - \psi_{AB}  + (4x_{(A}  -  \psi_{(A} )  (- 3x_{B)} + \psi_{B)})   - 3x_A x_B  + \psi_A \psi_{B}  \]
\[ +   (3x^C - \psi^C)  \Gamma_{(AB)C}  -    x^C\Gamma_{C(AB)}\]
\[ =   3x_{AB} - \psi_{AB} + (4x_{(A}  -  \psi_{(A} )  (- 3x_{B)} + \psi_{B)})   - 3x_A x_B  + \psi_A \psi_{B}   \]
\[ + (3x_{(A}  - \psi_{(A} ) (2x_{B)}  - \psi_{B)})   +    x_{(A}  ( 3x_{B)}  - \psi_{B)})\]
\[ =   3x_{AB} - \psi_{AB} - 6x_{A}x_B + x _{(A}\psi_{B)} + \psi_A\psi_B \]
\[ =   3x_{AB} - \psi_{AB}   + (\psi_{(A} - 2 x_{(A})(\psi_{B)} + 3x_{B)}).\]
So we have:
\[ Q_{(AB)} =  x\psi^{-1}\left(  - \pi^{A'} \pi^{B'}S_{ab} + 3x_{AB} - \psi_{AB}   + (\psi_{(A} - 2 x_{(A})(\psi_{B)} + 3x_{B)})\right).\]
For the metric we need the combination: $\displaystyle{U_{AB} = \frac{1}{5}(7P_{(AB)} - 2Q_{(AB)})}$.  We find:
\[ 10\psi x^{-1} U_{(AB)} - 4 \pi^{A'} \pi^{B'}S_{ab} \]\[ =  7\psi_{AB} -   28x_{AB}   - 21(\psi_{A}- 2x_A )(\psi_{B} - 2x_B )  - 12x_{AB}  + 4\psi_{AB}   - 4(\psi_{(A} - 2 x_{(A} )(\psi_{B)} + 3x_{B)} )\]
\[ =  11\psi_{AB}-   40x_{AB}    - 5(\psi_{(A} - 2x_{(A})(5\psi_{B)} - 6x_{B)}).\]
So we get:
\[ U_{AB} = \frac{x}{2\psi}\left(  \frac{4}{5} \pi^{A'} \pi^{B'}S_{ab} + \frac{11}{5}\psi_{AB} -   8x_{AB}  - (\psi_{(A} - 2x_{(A})(5\psi_{B)} - 6x_{B)})\right)\]
\[ = \frac{x}{10\psi}\left(  4 \pi^{A'} \pi^{B'}S_{ab} + 11\psi_{AB} - 25\psi_A \psi_B\right) + x\psi^{-1} \left( -   4x_{AB} + 8 \psi_{(A}x_{B)}  - 6x_Ax_B)\right).\]
\eject\noindent
\section{The $\mathbb{G}_2$-conformal structure}
The metric, regarded as a quadratic form on a tangent vector $Z = z^a \partial_a + z_{A'}\partial^{A'}$, is:
\[ g(Z, Z) = - \iota_{Z_3}\iota_{Z_1}(d\gamma) + \frac{2}{3}(\iota_Z (\gamma))^2, \]
\[ Z_3 = \pi_{-1}(Z) = Z_2 + \frac{1}{5}(7\Psi - 2\Phi)(Z_1). \]
Now we have:
\[ \iota_Z(\gamma) = x^{-1}(\pi^{C'}d\pi_{C'} + (3 x_A - \psi_A ) \pi_{A'} \theta^a)(z^b \partial_b + z_{B'}\partial^{B'})\]
\[ = x^{-1}\pi^{C'}z_{C'} + x^{-1}(3 x_A  - \psi_A ) \pi_{A'} z^a\]
\[ = x^{-1}\pi^{C'}z_{C'} - 3y^Ax_A + y^A\psi_A.\]
Here $y^A = -x^{-1} z^a\pi_{A'}$, as usual.  Note that $Z_1 = y^A D_A$.  Also $Z_3 = v^A D_A$, where we have:
\[ v^A =  u^A +  \frac{1}{5}((7\Psi- 2\Phi)(y))^A.\] Collecting terms in $\displaystyle{\frac{2}{3}(\iota_Z(\gamma))^2}$, we get:
\[ (\iota_Z(\gamma))^2 = F^{A'B}z_{A'}z_{B'} + F^a y_A z_{A'} +   F_{AB}y^A y^B, \]
\[ F^{A'B'} = \frac{2}{3}x^{-2} \pi^{A'} \pi^{B'}, \]
\[ F^a =   4x^{-1} \pi^{A'}x^A - \frac{4}{3}x^{-1} \pi^{A'}\psi^A, \]
\[ F_{AB} = 6x_Ax_B - 4x_{(A} \psi_{B)} + \frac{2}{3}\psi_A \psi_B.\]
It remains to compute the term  $\iota_{Z_3}\iota_{Z_1}(d\gamma)$.  First we have:
\[ d\gamma =  d(x^{-1}\pi^{C'}d\pi_{C'} + (3 x^{-2} D_A x - x^{-1}\psi^{-1} D_A \psi) \pi_{A'} \theta^a)\]
\[ = \pi^{C'} d\pi_{C'} \theta^a (x^{-2}\partial_{a}x + \partial_{A'} (3 x^{-2} D_A x - x^{-1}\psi^{-1} D_A \psi) ) \]    \[ +\theta^a\theta^b( (2x)^{-1}R_{abC'D'}\pi^{C'}\pi^{D'}   - \pi_{A'}\partial_b(3 x^{-2} D_A x - x^{-1}\psi^{-1} D_A \psi) ). \]
\eject\noindent
Contracting this expression with $D_A$, we get:
\[ \iota_{D_A}(d\gamma)  = \iota_{D_A}(d\gamma)\]
\[ = - \pi^{C'} d\pi_{C'} \pi^{A'} (x^{-2}\partial_{a}x + \partial_{A'} (3 x^{-2} D_A x - x^{-1}\psi^{-1} D_A \psi)) \]    
\[ +\pi^{A'}\theta^b( x^{-1}R_{abC'D'}\pi^{C'}\pi^{D'}  + \pi_{B'}\partial_a(3 x^{-2} D_B x - x^{-1}\psi^{-1} D_B\psi) \]
\[ = - x^{-1}\pi^{C'} d\pi_{C'}  (4 x_A  - \psi_A) \]    
\[ + \theta^b( x^{-1}R_{abC'D'}\pi^{C'}\pi^{D'} \pi^{A'} + \pi_{B'}D_A(3 x^{-2} D_B x - x^{-1}\psi^{-1} D_B\psi) \]
\[ = - x^{-1}\pi^{C'} d\pi_{C'}  (4 x_A  - \psi_A) \]    
\[ + \theta^b( x^{-1}(   - \epsilon_{AB}\psi_{B'}-  \pi_{B'}S_{ABC'D'}\pi^{C'}\pi^{D'} )+ \pi_{B'}D_A(3 x^{-2} D_B x - x^{-1}\psi^{-1} D_B\psi)) \]
\[ = - x^{-1}\pi^{C'} d\pi_{C'}  (4 x_A  - \psi_A)  - x^{-1}\theta_a\psi^{A'}-   x^{-1}\theta^BS_{ab}\pi^{A'}\pi^{B'} + \theta^B D_A(3 x^{-2} D_B x - x^{-1}\psi^{-1} D_B\psi). \]
Then we have:
\[ \iota_{D_B} \iota_{D_A}(d\gamma) = - \epsilon_{AB} x^{-1}\psi, \hspace{10pt}  \iota_{Z_3}\iota_{Z_1}(d\gamma)  = - x^{-1} \psi y_A v^A.\]
Thus it remains to compute the term  $x^{-1} \psi y_A v^A$.
Now we have:
\[ y_Av^A =  y_Au^A  +  \frac{y_A}{5}((7\Psi - 2\Phi)(y))^A\]
\[ = y_Au^A  +  y_AX(y^A) - \frac{1}{5}(7P_{AB}- 2Q_{AB})y^Ay^B\]
\[ = y_A(x^{-1}z^{a}x_{A'} - X( y^A) + y^BD_B \xi^A  - k\xi^A) + y_AX(y^A) - U_{AB}y^Ay^B\]
\[ = x^{-1}y_Az^{a}x_{A'} - y^A y^B( D_A \xi_B + U_{AB}) + k\xi_Ay^A\]
\[ = -x^{-2}z_A^{B'} \pi_{B'} z^{AA'} x_{A'}  - y^A y^B( D_A \xi_B + U_{AB}) 
 \] 
 \[ + (\xi_Ay^A)(x^{-1}z_{A'}\pi^{A'} + y^B (\psi_B - 3x_B))\]
\[ = \frac{1}{2x}z_az^a  - y^A y^B( D_A \xi_B + U_{AB})  + (\xi_Ay^A)(x^{-1}z_{A'}\pi^{A'} + y^B (\psi_B  - 3x_B))\]
\[ = (2x)^{-1}z_az^a  - x^{-1}y_A z_{A'}\xi^A\pi^{A'} \]
\[ + y^Ay^B( - D_A \xi_B - U_{AB} + \xi_A (\psi_B  - 3x_B )).\]
\eject\noindent
So  we get:
\[ x^{-1} \psi y_A v^A = H_{ab} z^a z^b + H^a y_A z_{A'} + y^Ay^B H_{AB}, \]
\[ H_{ab} = \frac{\psi}{2x^2} g_{ab}, \]
\[ H^a = -x^{-1} \pi^{A'}(4x^A - \psi^A), \]
\[ H_{AB} = x^{-1} \psi ( - D_{(A} \xi_{B)} - \xi_{(A}  (3x_{B)} - \psi_{B)} ) - U_{AB})\]
\[ =  - x^{-1} \psi D_{(A} (4\psi^{-1} D_{B)}x - x\psi^{-2}D_{B)}\psi) - (4x_{(A}- \psi_{(A}) (3x_{B)} - \psi_{B)} ) -   x^{-1} \psi U_{AB}\]
\[ =\psi_{AB}  - 4x_{AB}  + 5\psi_{(A}x_{B)} -2\psi_A\psi_B   - (4x_{(A}- \psi_{(A}) (3x_{B)} - \psi_{B)} ) -   x^{-1} \psi U_{AB}\] 
\[ =\psi_{AB}  - 4x_{AB}  - 12x_Ax_B + 12\psi_{(A}x_{B)} -3\psi_A\psi_B  -   x^{-1} \psi U_{AB}\] 
\[ =\psi_{AB}  - 4x_{AB}  -  3(\psi_{(A} - 2x_{(A})(\psi_{B)} - 2 x_{B)})\]\[  -   \frac{2}{5} \pi^{A'} \pi^{B'}S_{ab} - \frac{11}{10}\psi_{AB} + \frac{5}{2}\psi_A \psi_B +    4x_{AB} - 8 \psi_{(A}x_{B)}  + 6x_Ax_B\]
\[ =  -  \frac{2}{5} \pi^{A'} \pi^{B'}S_{ab}  - \frac{1}{10}\psi_{AB}   - \frac{1}{2}\psi_A \psi_B  + 4 \psi_{(A}x_{B)}  - 6x_Ax_B.\]
Note that we have: 
\[   F^a + H^a = 4x^{-1} \pi^{A'}x^A - \frac{4}{3}x^{-1} \pi^{A'}\psi^A -x^{-1} \pi^{A'}(4x^A - \psi^A)  =  - \frac{1}{3}x^{-1} \pi^{A'}\psi^A , \]
\[ F_{AB} + H_{AB} = 6x_Ax_B - 4x_{(A} \psi_{B)} + \frac{2}{3}\psi_A \psi_B  -  \frac{2}{5} \pi^{A'} \pi^{B'}S_{ab}  - \frac{1}{10}\psi_{AB}   - \frac{1}{2}\psi_A \psi_B  + 4 \psi_{(A}x_{B)}  - 6x_Ax_B\]
\[ =  \frac{1}{6}\psi_A \psi_B  -  \frac{2}{5} \pi^{A'} \pi^{B'}S_{ab}  - \frac{1}{10}\psi_{AB} . \]
So now we may collect terms in the metric:
\[ g(Z, Z) = F^{A'B'}z_{A'}z_{B'} + K^a y_{A} z_{A'} + H_{ab} z^a z^b + y^Ay^B G_{AB}. \]
\[ F^{A'B'} = \frac{2}{3x^2} \pi^{A'}\pi^{B'}, \]
\[ K^a = F^a + H^a =  - \frac{1}{3x} \pi^{A'}\psi^A , \]
\[ H_{ab} = \frac{\psi}{2x^2} g_{ab}, \]
\[  G_{AB} =  F_{AB} + H_{AB} =   - \frac{2}{5} \pi^{A'} \pi^{B'}S_{ab}+ \frac{1}{6}\psi_A \psi_B    - \frac{1}{10}\psi_{AB}.\]
Alternatively, after substituting for $y^A$, we have:
\[ x^2g(Z, Z) = G^{A'B'}z_{A'}z_{B'} +  2G_{a}^{B'} z^az_{B'} + G_{ab} z^a z^b, \]
\[ G^{A'B'} = \frac{2}{3} \pi^{A'}\pi^{B'}, \hspace{10pt} G_a^{B'} = - \frac{1}{6} \psi_{A}\pi_{A'} \pi^{B'}, \]
\[ G_{ab} =  \frac{\psi}{2} g_{ab} + \frac{1}{30}\pi_{A'} \pi_{B'}(- 12 \pi^{C'} \pi^{D'}S_{ABC'D'}+ 5\psi_A \psi_B    - 3\psi_{AB}).\]
Remarkably, all the dependence of the metric on the vector field $X$ goes only into the conformal factor $x^2$!
\\\\
Writing the metric in terms of co-vectors, we have:
\[ g = F^{A'B'}d\pi_{A'}\otimes d\pi_{B'} - \frac{1}{2}x^{-1}K^a (\theta_{A}\otimes d\pi_{A'} + d\pi_{A'}\otimes \theta_A) + H_{ab} \theta^a\otimes  \theta^b + x^{-2}\theta^A\otimes \theta^B G_{AB} \]
\[ = \frac{2}{3x^2} (\pi^{A'}d\pi_{A'})^2 - \frac{1}{3x^2} \pi^{A'}d\pi_{A'}\psi_A \theta^{A} +  \frac{\psi}{2x^2} \theta^a  \theta_a \]\[+ x^{-2}\theta^A \theta^B \left( - \frac{2}{5} \pi^{A'} \pi^{B'}S_{ab}+ \frac{1}{6}\psi_A \psi_B    - \frac{1}{10}\psi_{AB}\right). \]
Rescaling,  we have:
\[ 120x^2 g =  80(\pi^{A'}d\pi_{A'})^2  - 40\pi^{A'}d\pi_{A'} \psi_A\theta^A  + 60\psi  \theta^a\theta_a + \chi_{AB} \theta^A \theta^B, \]
\[ \chi_{AB} =   - 48 \pi^{C'} \pi^{D'}S_{ABC'D'}+ 20\psi_A \psi_B    - 12\psi_{AB}.\]
Alternatively, completing the square with the $\pi^{A'} d\pi_{A'}$ terms, we get:
\[ 120x^2 g =   5 \left(4\pi^{A'}d\pi_{A'}  - \psi_A\theta^A \right)^2   + 60\psi \theta^a\theta_a + \gamma_{AB} \theta^A \theta^B, \]
\[ \gamma_{AB} =   - 48 \pi^{C'} \pi^{D'}S_{ABC'D'}+ 15 \psi_A \psi_B    - 12\psi_{AB}.\]
Rescaling again, we obtain the conformal structure in the form:
\[ G = 2x^2 \psi^{-1} g = \frac{1}{12 \psi} \left(4\pi^{A'}d\pi_{A'}  - \psi_A\theta^A \right)^2   +  \theta^a\theta_a + 2\theta^A \theta^B\tau_{AB}, \]
\[ \tau_{AB} =   \frac{1}{40\psi}(-16 \pi^{C'} \pi^{D'}S_{ABC'D'}+ 5 \psi_A \psi_B  - 4\psi_{AB}) = \tau_{BA}.\]
This completes the proof of theorem two.
\section{The Kapadia family of plane wave space-times}
The Kapadia family of space-times is a family of null plane-fronted waves, whose metric in co-ordinates $(u, v, x,y) \in \mathbb{U} \subset \mathbb{R}^4$ (with $\mathbb{U}$ open and connected, for simplicity) is \cite{kap}:
\[ g = 2h^{-2}(u)(dudv  - dx^2 - f^2(u)dy^2) = 2(ln - m\overline{m}), \]
\[  l = h^{-2}du, \hspace{10pt} n = dv, \hspace{10pt} m = h^{-1}(dx + ifdy),\hspace{10pt} \overline{m} = h^{-1}(dx -
ifdy), \]
\[ dl = 0, \hspace{12pt} dn = 0, \hspace{12pt} dm = Cl, \hspace{12pt} d\overline{m} = \overline{C}l,\]
\[ C = Am +B\overline{m} = h'dx + i(fh' - hf')dy, \]
\[ dC = Slm + Tl\overline{m}, \]
\[ A = \frac{h}{2f}(2fh' -f'h), \hspace{10pt} B = \frac{h^2f'}{2f}, \]
\[ S = h^2A' - A^2 - B^2 = h^4\left(\frac{h''}{h}- \frac{f''}{2f}\right), \hspace{10pt} T = h^2B' - 2AB = \frac{h^4f''}{2f}.\] 
Here $f(u)$ and $h(u)$ are positive smooth real functions of the variable $u$, defined on the open set $\mathbb{U}$.  Also we use a prime to abbreviate the derivative with respect to $u$.  For the spin connection, we take a spin basis, $\{o_A, \iota_A\}$, with conjugate basis,
$\{o_{A'}, \iota_{A'}\}$, such that  $l_a = o_Ao_{A'}$, $m_a = o_A\iota_{A'}$, $\overline{m}_a = \iota_{A}o_{A'}$
and $n_a = \iota_{A}\iota_{A'}$,   $2o_{[A'}\iota_{B']} = \epsilon_{A'B'}$ and $g_{ab} = \epsilon_{AB}\epsilon_{A'B'}$.  The associated Levi-Civita spin connection is simply:
\[ do_A = 0, \hspace{10pt} d\iota_A = \overline{C}o_A, \hspace{10pt}do_{A'} = 0, \hspace{10pt} d\iota_{A'} = Co_{A'}.\]
The primed curvature spinor, $R_{A'B'}$, given by the formula, valid for any spinor field $v_{A'}$:  $d^2v_{A'} = R_{A'B'}v^{B'}$, is as follows:
\[ R_{A'B'} = o_{A'}o_{B'}dC = - \Sigma^{C'D'}C_{A'B'C'D'} - S_{ab}\Sigma^{A'B'}, \]
\[ C_{A'B'C'D'} = - To_{A'}o_{B'}o_{C'}o_{D'}, \hspace{10pt} S_{ab} = Sl_al_b = So_Ao_Bo_{A'}o_{B'}.\]
In particular the Ricci scalar is zero; whenever  $f''$ is non-zero, the Weyl spinor $C_{A'B'C'D'}$ is non-zero and of type $(4)$ (null). Hence-forth we assume that $f''\ne 0$.  Note that the Kapadia space-times are always locally conformal to vacuum, since locally the equation $S =0$ can be solved with a positive solution for $h$, given $f$.
\eject\noindent
\section{The (2, 3, 5)-structure for the Kapadia metrics}
Introduce the fiber co-ordinate for the primed co-spin bundle,  $\lambda$, by the relation $0  = \pi_{A'} o^{A'}\lambda + \pi_{A'} \iota^{A'}$.  Then the torsion-free Weyl connection is: 
\[ \mathcal{T'} = \{ \alpha, \beta, \gamma\}, \]
\[ \alpha  = \lambda l + m = - \frac{o_A \pi_{A'} \theta^a}{\pi_{B'}o^{B'}}, \hspace{10pt} \beta = \lambda \overline{m} + n =  - \frac{\iota_A \pi_{A'} \theta^a}{\pi_{B'}o^{B'}}, \hspace{10pt}  \gamma = d\lambda + C =   \frac{ \pi^{A'}d\pi_{A'}}{(\pi_{B'}o^{B'})^2}.\]
Here we require that $t \ne 0$ and we have put $\displaystyle{\lambda = st^{-1}}$.  Note that the  the principal spinor of the Weyl spinor corresponds to $\lambda = \infty$, so the Weyl structure is defined for all complex $\lambda$.  By straight-forward manipulations, we may put this structure in the standard Cartan form, with variables $\{u, t, p, q, z\}$:
\[ \mathcal{T}' = \{ dt - pdu,  dp - qdu, dz  - Fdu\}.\]
Here we have:
\[ ht = - x - iyf, \]
\[ h^2p = \lambda + xh' + iy(fh' -hf'), \]
\[ h^3q =  - 2h' (\lambda) +x (hh''- 2(h')^2) + iyf(hh''- 2(h')^2) - iyh(hf'' - 2h'f'),   \]
\[ z = v + (\lambda h^{-1})(x - ify) + \frac{f'}{4f}(x - ify)^2, \]
\[ F  = \alpha q^2 + \beta pt + \gamma qt + \delta t^2 + \epsilon pq + \zeta p^2, \]
\[ \alpha = \frac{fh^2}{f''}, \hspace{10pt}  \beta =  - 2hh'+ \frac{h^2f'}{2f} + \frac{4h'h''f}{f''}, \hspace{10pt} \gamma = \frac{2hh''f}{f''} - h^2,\]\[  \delta =  \frac{h^2f''}{4f}  + \frac{(h'')^2f}{f''}  - hh''  - \frac{h^2(f')^2}{4f^2}, \hspace{10pt} \epsilon =  \frac{4fhh'}{f''}, \hspace{10pt} \zeta = \frac{4f(h')^2}{f''}.    \]
So we have shown that the Cartan function $F(u, t, p, q, z)$ for the Kapadia space-times can be taken to be a homogeneous  quadratic function of the variable $t$, $p$ and $q$, with coefficients functions of the variable $u$ only, determined directly from the metric functions $f(u)$ and $h(u)$.  Note that, in particular, the function $F$ is independent of the variable $z$.  Also we have $F_{qq} = 2\alpha \ne 0$, so the Cartan formalism applies.  
\section{The $\mathbb{G}_2$-structure for the Kapadia metrics}
Using Maple and the metric formula of Nurowski, the conformal structure $G$ for the Weyl structure of the spin bundle for the Kapadia space-times takes the form:
\[G = G_{11}(dt - pdu)^2 + G_{33}(dp - qdu)^2 \]\[+ 2(dt - pdu)(G_{13}(dp - qdu)+G_{14}dq+G_{15} du)+2G_{25} du ( dz - F du - F_q(dp - qdu)).\]
Here the not identically zero pentad coefficients are  given as follows:
\[ \frac{(f'')^6}{16h^6 f^2} G_{11} = h^2(f')^2 (f'')^2 +10hh'ff'(f'')^2+4h^2ff'f''f'''-20(h')^2f^2(f'')^2\]\[ -10hh'f^2f''f''' -5h^2f^2(f''')^2-12h^2f(f'')^3
+30hh''f^2(f'')^2+3f^2h^2f''f'''',\]
\[ \frac{(f'')^5}{80h^7 f^3} G_{13} =   hf'f'' +2h'ff''-hff''',\]\[ G_{14} =  \frac{240 f^4 h^8}{(f'')^4}, \hspace{10pt} G_{25} =  \frac{120 f^3 h^6}{(f'')^3}, \hspace{10pt} G_{33} = - \frac{320f^4 h^8}{(f'')^4},  \]
\[  \frac{(f'')^5}{60f^2h^6} G_{15} =  4qhf(hf'f''+ 2h'ff''- hff''')\]
\[ + t(- 4h'h''f^2f'' +  4hh'''f^2f''+ 4hh''ff'f''- 4hh''f^2f''' - h^2f'(f'')^2)\]
\[ + 2pf(4hh'f'f'' - 4(h')^2ff''+ 6hh''ff''- 4hh'ff'''- h^2(f'')^2). \]
Maple gives the determinant of the metric in the co-ordinates $(u, t, p, q, z)$ as follows: $- 2^{20} 3^4 5^5f^{18} h^{36} (f'')^{-18}$, which is non-zero.
The Einstein and Weyl tensors of the metric $G$, as just written, each have only one independent component.  These components may be written quite compactly by introducing functions $t(u)$ and $w(u) > 0$, related to $f$ and $h$ by the formulas:
\[ e^{\int s(u) du} = \frac{f''}{f}, \hspace{10pt} h^2 = \left(\frac{f''}{f}\right)w^{-\frac{2}{3}}, \hspace{10pt} s = \frac{f'''}{f''} - \frac{f'}{f}, \hspace{10pt}\epsilon w^2 =  \left(\frac{f''}{fh^2}\right)^{3}, \hspace{10pt} \epsilon = \pm 1.\]
Then we find, using a co-ordinate basis in the order $(u, t, p, q, z)$ that the independent non-identically zero curvature components, $ET_{11}$ of the Einstein tensor and $WT_{1212}$ of the Weyl tensor are as follows:
\[ 20 w ET_{11} =  -60w'' + 3(s^2+ 2e^{\int s(u) du}- 4s')w, \]
\[5w^{\frac{8}{3}}WT_{1212} =   90ss'' - 96s^2s'+e^{\int s(u) du}(48 s'-12s^2) -27e^{2\int s(u) du}+57(s')^2+12s^4-30s'''. \]
It is easily seen that the metric $G$ is always locally conformal to vacuum.  Finally, translating back into the original co-ordinates of the spin bundle, using Maple we find exact agreement with the general formula given in Theorem 2.
\eject\noindent
\section{Special cases of the Kapadia $\mathbb{G}_2$-structures}
Consider the case that $f = u^{\frac{m+1}{2}}$, with $u> 0$ and $m \ne \pm 1$ a real constant.\\  Then we have, after multiplying the conformal structure by the conformal factor $w^2V^{-2}$:
\[ e^{\int s(u) du} =  \frac{m^2 -1}{4u^2}, \hspace{10pt} s = \frac{f'''}{f''} - \frac{f'}{f} = - \frac{2}{u}.\]
\[ 40 V ET_{11} =  -120V'' + 3(m^2 - 9)u^{-2}V, \]
\[80w^{\frac{2}{3}}V^2WT_{1212}  = - 3u^{-4}(m - 3)(m + 3)(3m - 1)(3m + 1).\]
In particular in the cases $f = u^2$, $f = u^{-1}$, $\displaystyle{f = u^{\frac{2}{3}}}$ and $\displaystyle{f = u^{\frac{1}{3}}}$, the conformal structure of the co-spin bundle is conformally flat and in all other cases it is not conformally flat.  Also in the cases $f = u^2$ and $f = u^{-1}$, the $\mathbb{G}_2$-structure is flat, with $V = 1$,  whereas in the cases  $\displaystyle{f = u^{\frac{2}{3}}}$ and $\displaystyle{f = u^{\frac{1}{3}}}$, the $\mathbb{G}_2$-structure is flat, with $\displaystyle{V = u^{\frac{1}{3}}}$.  Note that in all these cases the original space-time metric is not conformally flat.  We summarize with a Lemma:
\begin{lemma}  Consider the special Kapadia conformal structures:\[ ds^2 = 2(du dv - dx^2 - u^{m+1} dy^2).\]  Then the $\mathbb{G}_2$-conformal structure of the primed co-spin bundle is conformally flat if and only if $m = \pm 3$ or $m = \pm \frac{1}{3}$ and in these cases gives local flat models for the Cartan $(2, 3, 5)$-system: i.e. the Cartan structure has the Lie algebra of $\mathbb{G}_2$ as symmetries.  In particular,  for these special values of $m$, the primed co-spin bundle is locally conformal to a de-Sitter space in five dimensions.
\end{lemma}

\eject\noindent
\section{The Schwarzschild metric}
Let $m$ be a positive real constant, with the same dimensions as length.  Then the Schwarzschild metric is:
\[ ds^2 = \left(\frac{r - 2m}{r}\right)\left(dt^2-  \frac{r^2dr^2} {\left(r- 2m\right)^2}\right) - r^2 (d\theta^2 + \sin^2(\theta) d\phi^2)  = ln - \mu\mu',\]
\[ l = du = dt - \frac{rdr}{r - 2m}, \hspace{10pt} u = t - r - 2m\ln\left(\frac{r}{2m} - 1\right), \hspace{10pt} dl = 0, \]
\[ n = 2dr + \frac{(r - 2m)du}{ r},  \hspace{10pt}   dn = \frac{2m  dr du }{r^2}  =  \frac{m nl}{r^2}, \]
\[ \mu = r (d\theta + i \sin(\theta) d\phi) = r\sin(\theta) dx, \hspace{10pt} x =  i \phi +  \ln\left(\tan\left(\frac{\theta}{2}\right)\right), \]
\[ d\mu =  \frac{dr \mu }{r} + ri\cos(\theta) d \theta d\phi  =  \frac{r n\mu  +  (r - 2m)\mu l }{2r^2} -  \frac{1}{2r}  \cot(\theta) \mu \overline{\mu}.\]
Note that in the $(u, r, \theta, \phi)$ co-ordinate system the tetrad is defined for all $u$ and for $r > 0$. The variables $\theta$ and $\phi$ are standard spherical polar co-ordinates. The variable $x$ gives a local holomorphic co-ordinate for the natural complex structure of the two-spheres of constant curvature at constant $(t, r)$.  We introduce the spinor basis $o_{A'}$ and $\iota_{A'}$ with their conjugates $o_A$ and $\iota_A$, such that:
\[ l_a = o_{A} o_{A'}, \hspace{10pt} \mu_a = o_{A} \iota_{A'}, \hspace{10pt} \overline{\mu}_a = \iota_{A} o_{A'}, \hspace{10pt} n_a = \iota_{A} \iota_{A'}.\]
Then the Levi-Civita spin connection is:
\[ do_{A'} = Po_{A'} + B\iota_{A'}, \hspace{10pt} do_A = \overline{P} o_A + \overline{B} \iota_A, \]
\[ d\iota_{A'} = Do_{A'} - P\iota_{A'}, \hspace{10pt} d\iota_{A} = \overline{D} o_A - \overline{P} \iota_A, \] 
\[ P =  \left(\frac{m}{2r^2}\right) l + \left( \frac{1}{4r} \right) \cot(\theta) (\mu - \overline{\mu}) = \left(\frac{m}{2r^2}\right) du  + \frac{ i}{2} \cos(\theta) d\phi , \]
\[  B = -  \left(\frac{1}{2r}\right) \overline{\mu} = - \frac{1}{2} (d\theta - i \sin(\theta) d\phi),\]
\[ D = \left(\frac{r - 2m}{2r^2}\right) \mu =  \left(\frac{r - 2m}{2r}\right) (d\theta + i \sin(\theta) d\phi).\]
Then we have, for the curvature, valid for any spinor field $v_{A'}$:
\[ d^2 v_{A'} = R_{A'B'}v^{B'}, \hspace{10pt} R_{A'B'} = - \Sigma^{C'D'}C_{A'B'C'D'}, \hspace{10pt}  C_{A'B'C'D'} = \frac{3m}{r^3} o_{(A'} o_{B'} \iota_{C'} \iota_{D')}.\]
In particular, the space-time is Ricci flat: $S_{ab} = 0$ and $\Lambda =0$.  Also the self-dual Weyl curvature is never zero, except at the singularity $r = 0$.
\eject\noindent
\section{The $\mathbb{G}_2$-conformal structure of Schwarzschild}  In the primed co-spin bundle,  introduce the fiber co-ordinate $\lambda$ by the relation:  \[\pi_{A'} \iota^{A'} \lambda = \pi_{A'} o^{A'}.\]   Note that the principal spinors of the Weyl spinor, the zeroes of the polynomial $\psi$, correspond to the values $\lambda = 0$ and $\lambda = \infty$.  Using the formula of Theorem 2 above, a representative of the conformal structure $G$, which is defined whenever $\lambda \ne 0$ and $r \ne 0$, is given as follows:
\[ 12\psi (\pi_{B'} \iota^{B'})^{-4} G =  G_{11} (d\lambda)^2 + 2G_{12} d\lambda du + 2G_{13} d\lambda dr + 2G_{14} d\lambda dx + 2G_{15} d\lambda d\overline{x} \]\[ + G_{22} (du)^2 + 2G_{23} du dr + 2G_{24} du dx + 2G_{25} du d\overline{x}\]\[ + G_{33}(dr)^2 + 2G_{34} dr dx + 2G_{35} dr d\overline{x} +  G_{44}(dx)^2 + 2G_{45} dxd\overline{x} + G_{55} (d\overline{x})^2. \]
Here the individual coefficients are assembled as follows:
\[ G_{11} = 16, \]
\[ G_{12} = 4\lambda r^{-2}(5r- 14m), \]
\[ G_{13} = 20\lambda r^{-1}, \]
\[ G_{14} =   - 2\lambda^2 r^{-1} (r - 2m)\sin(\theta) - 8\lambda \cos(\theta), \]
\[ G_{15} =  8\lambda \cos(\theta) - 2\sin(\theta),  \]
\[ G_{22} = 4r^{-4}\lambda^2(4r^2 - 5rm  - 2m^2),\]
\[ G_{23} = 16\lambda^2 r^{-3}(r + 2m), \]
\[ G_{24} =  2\lambda^3 r^{-3}(r - 2m)(r - 4m)\sin(\theta)  - 2\lambda^2 r^{-2}(5r- 14m)  \cos(\theta),  \]
\[ G_{25} =   2\lambda^2 r^{-2} (5r- 14m) \cos(\theta)+  2\lambda r^{-2} (r - 4m)   \sin(\theta),    \]
\[ G_{33} = 28\lambda^2 r^{-2}, \] 
\[ G_{34} =  4\lambda^3 r^{-2}(2r - 7m)\sin(\theta) -   10\lambda^2 r^{-1} \cos(\theta), \]
\[ G_{35} =   10\lambda^2 r^{-1}  \cos(\theta)  - 4\lambda r^{-1} \sin(\theta),  \]
\[ G_{44} =  \lambda^4r^{-2}(r - 2m)^2\sin^2(\theta) +  2\lambda^3r^{-1} (r - 2m)\sin(\theta)\cos(\theta) + 4\lambda^2\cos^2(\theta), \]
\[ G_{45} =     - \lambda^3 r^{-1}(r - 2m) \sin(\theta) \cos(\theta)   - 5\lambda^2 r^{-1} (r + 4m)\sin^2(\theta)   - 4\lambda^2 \cos^2(\theta)  +\lambda   \sin(\theta) \cos(\theta),  \]
\[ G_{55} = 4\lambda^2 \cos^2(\theta) - 2\lambda\sin(\theta) \cos(\theta) + \sin^2(\theta).\]
Using Maple, the determinant of the matrix $\displaystyle{\frac{(\pi_{A'} \iota^{A'})^4 G_{ij}}{12\psi}}$ is  $\displaystyle{\frac{16r^{7}\sin^4(\theta)}{9m\lambda^2}}$.
It may be shown that this metric is regular on the axis, where $\theta = 0$, or $\theta = \pi$.
\eject\noindent
\section{The Cartan canonical form for Schwarzschild}
The torsion-free Weyl structure for Schwarzschild is as follows:
\[ \mathcal{T}' =  \{ \omega_1, \omega_2, \omega_3\},\] 
\[\omega_1 =  \frac{o_A \theta^A}{\pi_{C'} \iota^{C'}} = l - \lambda \mu = du - \lambda r\sin(\theta) dx,  \]
\[ \omega_2 = \frac{\iota_A \theta^A}{\pi_{C'} \iota^{C'}}  =  \overline{\mu} - \lambda n =  r\sin(\theta) d\overline{x} - \left(\frac{\lambda}{r}\right)(2rdr +  (r - 2m) du),    \]
\[ \omega_3 = \frac{\pi^{A'} d\pi_{A'}}{(\pi_{C'} \iota^{C'})^2}  =  d\lambda - B - 2\lambda P  +\lambda^2  D\]
\[ =  d\lambda -  \left(\frac{\lambda m}{r^2}\right) l  +  \left(\frac{\lambda}{2r^2\sin(\theta)}\right) (\lambda(r - 2m)\sin(\theta) - r\cos(\theta)) \mu + \left( \frac{1}{2r\sin(\theta)}\right)( \lambda \cos(\theta) + \sin(\theta)) \overline{\mu} \]
\[ =  d\lambda -  \left(\frac{\lambda m}{r^2}\right) du  +  \left(\frac{\lambda}{2r}\right) (\lambda(r - 2m)\sin(\theta) - r\cos(\theta)) dx + \frac{1}{2}( \lambda \cos(\theta) + \sin(\theta)) d\overline{x}. \]
Now put:
\[ p = \lambda r\sin(\theta), \hspace{10pt} q =  \lambda r\sin(\theta)\cos(\theta) - \lambda^2 \sin^2(\theta)(r - 3m), \hspace{10pt} z=  \cos(\theta) -  \lambda\sin(\theta), \]
Then we have:
\[ du - pdx = \omega_1, \]
\[ dp- qdx = -  \frac{ \lambda \sin(\theta)}{2r} (r - 4m) \omega_1  -  \frac{1}{2} \sin(\theta)\omega_2 + r\sin(\theta) \omega_3,\]
\[ dz - Fdx =  - \frac{m \lambda \sin(\theta)}{r^2} \omega_1-   \sin(\theta) \omega_3, \]
\[ - 2r^3F =  2qr^2 + r^3\sin^2(\theta)  + p^2(r - 2m).\]
The variables for the Cartan system are $u, p, q, z, x$.  In these variables, we have the Pfaffian system in the standard Cartan form:
\[ \mathcal{T}' = \{ du - pdx, \hspace{10pt} dp - qdx, \hspace{10pt} dz - F(x, u, p, q, z)dx\}, \]
\[ F =  \frac{z^2 -1}{2} - \frac{2( q- pz)^{\frac{3}{2}}}{3p(3m)^{\frac{1}{2}}}.\]
Note that $F$ is independent of the variables $x$ and $u$.  This reflects that the system is time independent and rotationally symmetric.  Calculating the conformal structure, directly in terms of $F$, using the Maple computing system and the metric formula of Nurowski, we find exact agreement with our general formula for the conformal structure.
\eject

\noindent
\begin{thebibliography}{20}
\bibitem{agr} Ilka Agricola, \\\textit{Old and New on the Exceptional Group $\mathbb{G}_2$}, \\
Notices of the American Mathematical Society, \textbf{55}, 922-929, 2008.
\bibitem{bae}John Baez, \\\textit{The octonions}, \\Bulletin of the American Mathematical Society, \textbf{39}, 145-205, 2002;  \\errata: Bulletin of the American Mathematical Society, \textbf{42}, 213, 2005.
\bibitem{cap}Andreas \v{C}ap and Katja Sagerschnig,\\ \textit{On Nurowski's conformal structure associated to a generic rank two distribution in dimension five}, \\October 2007, 23 pages,  arXiv:0710.2208
\bibitem{car1} \'{E}lie Cartan, \\\textit{Sur la structure des groupes de transformations finis et continues},\\ Th\`{e}se, Paris, Nony, 1894.
\bibitem{car2}\'{E}lie Cartan, \\\textit{Les syst\`{e}mes de Pfaff \`{a} cinq variables et les \'{e}quations aux deriv\'{e}es
partielles du second ordre}, \\Annales Scientifiques de l'\'{E}cole Normale Sup\'{e}rieure, \textbf{27}, 109-192, 1910.
\bibitem{dou1} Boris Doubrov and Igor Zelenko, 
\\\textit{A Canonical Frame for Nonholonomic Rank Two Distributions of Maximal Class}, \\
Comptes Rendus de l'Acad\'{e}mie des sciences, Paris, S\'{e}rie I, \textbf{342}, 589-594, 2006.
\bibitem{dou2} Boris Doubrov and Igor Zelenko, 
\\\textit{On local geometry of nonholonomic rank 2 distributions}\\
March 2007, 21 pages, arXiv:math/0703662, 2007.
\bibitem{eng}  Friedrich Engel, \\\textit{Ein neues, dem linearen Komplexe
analoges Gebilde},\\ Leipziger Berichte,  \textbf{52}, 63-76,
220-239, 1900.
\bibitem{fef1} Charles Fefferman and C. Robin Graham, \textit{Conformal invariants}, \\in "\'{E}lie Cartan et les
Math\'{e}matiques d'aujourd'hui," Ast\'{e}risque, Numero hors S\'{e}rie, 95-116, 1985.
\bibitem{fef2} Charles Fefferman and C. Robin Graham, \textit{The ambient metric}, \\
October 2008, 100 pages, arXiv:0710.0919v2, 2008.
\bibitem{hil} David Hilbert,\\ \textit{\"{U}ber den Begriff der Klasse von Differentialgleichungen}, \\Mathematische Annalen, \textbf{73}, 95-108, 1912.
\bibitem{kap} Devendra Kapadia and George Sparling, \\\textit{ A class of conformally Einstein
metrics}, \\Classical and Quantum Gravity, \textbf{24}, 4765-4776, 2000.
\bibitem{kil} Wilhelm Killing, \\\textit{Die Zusammensetzung der stetigen endlichen Transformationsgruppen},\\Mathematische Annalen, \textbf{33}, 1-48, 1888.
\bibitem{mon} Gaspard Monge, \\\textit{De l'analyse \`{a}  la G\'{e}om\'{e}trie, Cinqui\`{e}me \'{E}dition, Revue, Corrig\'{e}e et Annot\'{e}e par M. Liouville}, \\Bachelier, Paris, 1850.
\bibitem{new} Ezra Newman, \\\textit{Heaven and its properties},\\ General Relativity and Gravitation,
\textbf{7}, 107-127, 1976.
\bibitem{nur1} Pawel Nurowski,\\\textit{Differential equations and conformal structures}, \\Journal of Geometry and  Physics, 
\textbf{55}, 19-49, 2005.
\bibitem{nur2} Pawel Nurowski, \\\textit{Conformal structures with explicit ambient metrics and conformal $\mathbb{G}_2$ holonomy}, \\To appear in "Proceedings of the Workshop at 2006 IMA Summer Program "Symmetries and Overdetermined Systems of Partial Differential Equations", Minneapolis, July 2006.
\eject\noindent
\bibitem{nur3} Pawel Nurowski and George Sparling, \\\textit{Three dimensional Cauchy-Riemann structures and second
order ordinary differential equations},\\ Classical and Quantum Gravity, \textbf{20}, 4995-5016, 2003.
\bibitem{pen0} Roger Penrose, \\\textit{Non-linear gravitons and curved twistor theory}, \\General Relativity
and Gravitation, \textbf{7}, 31-52, 1976.
\bibitem{pen1} Roger Penrose and Wolfgang Rindler,\\
\textit{ Spinors and Space-Time, Volume 1:
Two-spinor Calculus and Relativistic Fields}, \\Cambridge: Cambridge University
Press, 1984.
\bibitem{pen2}Roger Penrose and Wolfgang Rindler, \\\textit{Spinors and space-time, Volume 2, Spinor and twistor methods in space-time geometry}, \\  Cambridge Monographs on Mathematical Physics, Cambridge University Press, 1988.
\bibitem{schw} Karl Schwarzschild, \\
\textit{\"{U}ber das Gravitationsfeld eines Massenpunktes nach der Einstein'schen Theorie},\\ Sitzungsberichte der K\"{o}niglich Preussischen Akademie der Wissenschaften, \textbf{1}, 189-196, 1916.
\bibitem{spa1} George Sparling, \\\textit{The twistor theory of hypersurfaces in space-time}, \\in
"Further Advances in Twistor Theory, Volume III", editors Lionel Mason, Lane
Hughston, Piotr Kobak and Klaus Pulverer, London: Pitman Press, 2001.
\bibitem{spa2} George Sparling, \\\textit{Germ of a synthesis: spaceÐtime is spinorial, extra dimensions are time-like},\\
Proceedings of the Royal Society A: Mathematical, Physical and Engineering Sciences,
\textbf{463}(2083), 1665-1679, 2007.
\end{thebibliography}
\end{document}